\documentclass[twocolumn,times,tighten,astrosym]{aastex61}

\usepackage{graphicx, latexsym, graphicx, amssymb, longtable, epsf}
\usepackage{amssymb, amsmath, mathrsfs, subfigure, threeparttable}

\hypersetup{linkcolor=cyan, citecolor=cyan, filecolor=cyan, urlcolor=blue}

\newcommand{\myr}{${\rm M_{\sun}\,yr^{-1}}$}
\newcommand{\Msol}{${\rm M_{\sun}}$}

\newcommand{\um}{${\rm \mu}$m}

\newcommand{\Jykms}{${\rm Jy\,km\,s^{-1}}$}
\newcommand{\kms}{${\rm km}\,{\rm s}^{-1}$}

\newcommand{\alphaCO}{${\alpha_{\rm CO}}$}
\newcommand{\asec}{$^{\prime\prime}$}

\newcommand{\subfigsztwo}{1.05}

\received{2018}
\revised{}
\accepted{}

\shorttitle{Resolved CO (2--1) Observations in $z\sim1.6$ Cluster Galaxies}
\shortauthors{Noble et al.}

\begin{document}

\title{Resolving CO (2-1) in $\lowercase{z}\sim1.6$ Gas-Rich Cluster Galaxies with ALMA: Rotating Molecular Gas Disks with Possible Signatures of Gas Stripping}

\correspondingauthor{A.G.~Noble}
\email{noble@mit.edu}

\author{A.G.~Noble}
\affiliation{Kavli Institute for Astrophysics and Space Research, Massachusetts Institute of Technology, 77 Massachusetts Avenue, Cambridge, MA 02139, USA}

\author{A.~Muzzin}
\affiliation{Department of Physics and Astronomy, York University, 4700 Keele Street, Toronto, ON MJ3 1P3 Canada}

\author{M.~McDonald}
\affiliation{Kavli Institute for Astrophysics and Space Research, Massachusetts Institute of Technology, 77 Massachusetts Avenue, Cambridge, MA 02139, USA}

\author{G.~Rudnick}
\affiliation{The University of Kansas, Department of Physics and Astronomy, 1251 Wescoe Hall Drive, Lawrence, KS 66045, USA}

\author{J.~Matharu}
\affiliation{Institute of Astronomy, University of Cambridge, Madingley Road, Cambridge CB3 0HA, UK}

\author{M.C.~Cooper}
\affiliation{Department of Physics and Astronomy, University of California, Irvine, 4129 Frederick Reines Hall, Irvine, CA 92697, USA}

\author{R.~Demarco}
\affiliation{Departamento de Astronom\'{i}a, Facultad de Ciencias F\'{i}sicas y Matem\'{a}ticas, Universidad de Concepci\'{o}n, Concepci\'{o}n, Chile}

\author{C.~Lidman}
\affiliation{The Research School of Astronomy and Astrophysics, Australian National University, ACT 2601, Australia}

\author{J.~Nantais}
\affiliation{Departamento de Ciencias F\'{i}sicas, Universidad Andres Bello, Fernandez Concha 700, Las Condes 7591538, Santiago, Regi\'{o}n Metropolitana, Chile}

\author{E.~van Kampen}
\affiliation{European Southern Observatory, Karl-Schwarzschild-Strasse 2 D-85748 Garching bei M\"{u}nchen, Germany}

\author{T.M.A.~Webb}
\affiliation{Department of Physics, McGill University, 3600 rue University, Montr\'{e}al, QC H3A 2T8, Canada}

\author{G.~Wilson}
\affiliation{Department of Physics and Astronomy, University of California, Riverside, CA 92521, USA}

\author{H.K.C.~Yee}
\affiliation{Department of Astronomy and Astrophysics, University of Toronto, 50 St. George Street, Toronto, ON M5S 3H4, Canada}

\begin{abstract}\noindent

We present the first spatially-resolved observations of molecular gas in a sample of cluster galaxies beyond $z>0.1$.  Using ALMA, we detect CO (2--1) in eight $z\sim1.6$ cluster galaxies, all within a single 70\arcsec\ primary beam, in under three hours of integration time.  The cluster, SpARCS-J0225, was discovered by the Spitzer Adaptation of the Red-sequence Cluster Survey,  and is replete with gas-rich galaxies in close proximity.  It thus affords an efficient multiplexing strategy to build up the first sample of resolved CO in distant galaxy clusters. Mapping out the kinematic structure and morphology of the molecular gas on $\sim$3.5 kiloparsec scales reveals rotating gas disks in the majority of the galaxies, as evidenced by smooth velocity gradients.  Detailed velocity maps also uncover kinematic peculiarities, including a central gas void, a merger, and a few one-sided gas tails.   We compare the extent of the molecular gas component to that of the optical stellar component, measured with rest-frame optical \textit{HST} imaging.  We find that the cluster galaxies, while broadly consistent with a ratio of unity for stellar-to-gas effective radii, have a moderately larger ratio compared to the coeval field; this is consistent with the more pronounced trend in the low-redshift Universe.  Thus, at first glance, the $z\sim1.6$ cluster galaxies generally look like galaxies infalling from the field, with typical main-sequence star formation rates and massive molecular gas reservoirs situated in rotating disks.  However, there are potentially important differences from their field counterparts, including elevated gas fractions, slightly smaller CO disks, and possible asymmetric gas tails.  Taken in tandem, these signatures are tentative evidence for gas-stripping in the $z\sim1.6$ cluster.  However, the current sample size of spatially-resolved molecular gas in galaxies at high redshift is small, and verification of these trends will require much larger samples of both cluster and field galaxies.

\end{abstract}


\section{Introduction}

The advent of integral field spectroscopy combined with spatial multiplexing has enabled statistical kinematic studies of spatially-resolved ionized gas in hundreds of galaxies out to $z\sim3$ (\citealp[e.g.,][]{Forster09, Croom12, Wisnioski15, Bundy15, Forster18}). 
One of the major discoveries to stem from these data is the ubiquity of extended and rotationally-supported H$\rm{\alpha}$ gas disks at high redshift (\citealp[e.g.,][]{Forster06}), favoring a mode of galaxy growth through continuous star formation where gas is smoothly accreted via cold streams (\citealp[e.g.,][]{Keres05, Dekel09}).  This is further consistent with the existence of a tight main-sequence of star formation (\citealp[e.g.,][]{Noeske07}) and high integrated molecular gas fractions in $z>1$ galaxies (\citealp[e.g.,][]{Tacconi10, Daddi10, Genzel10, Papovich16, Noble17}).  

However, unlike the well-studied ionized gas component, spatially-resolved kinematic studies of the molecular gas component---the raw fuel for star formation---are still in their infancy.
The handful of spatially-resolved studies that do exist at high redshift have typically relied on extreme objects, such as submillimeter galaxies, quasars, and Brightest Cluster Galaxies (\citealp[e.g.,][]{Genzel03, Riechers08, Bothwell10, Hodge12, Russell17}), or rare, gravitationally-lensed objects (\citealp[e.g.,][]{Sharon13, Rawle14, Gonzalez17, Sharda18}), enabling detailed mapping of the molecular gas component.  Owing to the large time investment, the $z>1$ main-sequence population is instead beset with molecular gas observations that are marginally-resolved; fewer than ten detections with sub-arcsecond angular resolution exist \citep{Tacconi10, Tacconi13, Genzel13, Cibinel17, Herrera18} and an additional $\sim15$ that are only resolved on $>10$\,kpc scales \citep{Tacconi13, Daddi10}.  Whether the molecular gas component in typical main-sequence galaxies is similar to the ionized gas (as probed by H$\rm{\alpha}$ studies), in both spatial extent and velocity structure, thus largely remains unanswered at $z>1$.

We now have large samples of \textit{integrated} molecular gas on $\gtrsim$15\,kpc scales in $z>1$ main-sequence galaxies within field \citep{Tacconi13, Tacconi18} and cluster environments \citep{Hayashi17, Noble17, Rudnick17, Stach17, Coogan18, Castignani18}.  With the high spatial and velocity resolution afforded by ALMA, we can now target known gas-rich main-sequence galaxies to \textit{resolve} their molecular gas.  Moreover, $z>1.5$ galaxy clusters, with intrinsically high surface densities of star-forming galaxies \citep{Tran10, Brodwin13, Nantais17}, are the obvious laboratories within which to exploit this untapped potential, given the high return with multiplexing.

Here we present the first observations of spatially-resolved molecular gas in cluster galaxies beyond $z>0.1$.  Using ALMA Cycle 5 observations with $\sim0.4$\asec\ resolution, we detect CO (2--1) in eight cluster galaxies at $z\sim1.6$, producing exquisite velocity maps over $\sim3.5$\,kpc scales.  Due to the high cluster density and large ALMA primary beam of $\sim$70\asec\ (FWHM), all eight detections are within a single pointing with only 2.7 hours of integration. Throughout the analysis, we use a Chabrier initial mass function \citep{Chabrier03} for stellar masses and star formation rates, a Galactic \alphaCO\ of $4.36$ for CO-to-H$_2$ \citep{Bolatto13} which also includes a 36\% correction for Helium, and a $\Lambda$CDM cosmology with $\Omega_{{\rm M}}=0.3$, $\Omega_{\Lambda}$ =0.7, and ${\rm H_{0} = 70\,km\,s^{-1}\,Mpc^{-1}}$

\section{Observations and Analysis}

\subsection{A $z\sim1.6$ SpARCS Cluster}
SpARCS J022546-035517 (J0225, \citealp{Nantais16}) was identified in the 42 sq.\ deg.\ SpARCS fields \citep{Muzzin09, Wilson09, Demarco10} through the Stellar Bump Sequence technique, in which the rest-frame near-infrared stellar spectral feature at 1.6\um\ is redshifted into the Infrared Array Camera (IRAC) filters aboard \textit{Spitzer} for $1.3<z<1.8$ galaxies \citep{Papovich10, Muzzin13}.  The cluster redshift of $z=1.60$ is determined via $>20$ spectroscopically-identified cluster members. Ancillary imaging over 16 bands from optical/near-infrared ($ugrizYK$s and F160W) to infrared/far-infared (3.6/4.5/5.8/8.0/24/250/350/500\um) provides estimates of photometric redshifts, stellar masses, and star formation rates over the entire cluster field.  Imaging details and analysis for $u$-band to 8\um\ data, along with stellar masses, are presented in \cite{Nantais16}, with infrared-determined star formation rates described in Section 2.4 of \cite{Noble17}.  We note some values have changed slightly with the inclusion of \textit{HST} photometry, and updated 24\um\ priors used in the star formation rates.

\subsection{ALMA Observations and Maps}
In \cite{Noble17}, we presented ALMA Cycle 3 data of integrated CO (2--1) in multiple pointings over three $z\sim1.6$ galaxy clusters.  We found evidence for systematically enhanced molecular gas fractions in these cluster galaxies, compared to coeval field galaxies with similar stellar masses and star formation rates.  In ALMA Cycle 5, we chose to observe the most gas-rich of these detections with a longer baseline, in order to spatially-resolve the CO (2--1) emission.  The data presented here thus consists of a single Band 3 pointing at 88\,GHz, totaling 4 hours (2.7 hours on-source), with $\sim0.4$\asec\ angular resolution.  We used the 1.875 GHz baseband in the frequency division correlator mode to detect CO (2--1) at $z\sim1.6$.  While the new Cycle 5 data (field of view $\sim70\arcsec$ across) overlaps with $\sim80$\% of two Cycle 3 pointings on J0225, it is centered on the edge of the primary beams in the previous data, optimizing the sensitivity for the most gas-rich source, while also including an additional three known CO detections from \cite{Noble17} and eight known spectroscopically-confirmed cluster members.

We calibrated the data using the standard ALMA Science Pipeline within CASA \citep[version 5.1.1,][]{McMullin07} and imaged the data cube with 0.04\asec\ pixels and spectral resolutions of 50 and 100\,\kms.  To maximize S/N, we used an inverse-variance weighting function (i.e.\ natural) when imaging the visibilities, and performed minimal cleaning on bright sources, down to a threshold of $\sim3\sigma$.  The final continuum-subtracted and primary-beam-corrected map has a central rms of $\sim 0.1$\,mJy\,beam$^{-1}$ per 50\,\kms\ channel and a synthesized beam of 0.50\asec$\times$0.41\asec.

Integrated-intensity (i.e.,\ moment 0) maps of the CO emission are created by collapsing the cube over frequency channels that encompass significant emission ($\gtrsim2\sigma$) for each source.  We compute the local rms in a $\sim$5\asec$\times$10\asec\ annulus around each source on its respective integrated CO map.  For higher S$/$N sources, we generate intensity-weighted velocity (i.e.,\ moment 1) maps from the 50\,\kms\ data cube, using only pixels with values $>3\sigma$, estimated from the local rms averaged over the relevant channels.  For lower S$/$N sources, we create the moment 0 and moment 1 maps from the 100\,\kms\ data cube to increase their detection significance. 

\begin{deluxetable*}{lcccccccccc}
\tablecaption{Properties of the resolved CO-detected cluster galaxies}
\label{tab:results}
\tablehead{
\colhead{ID} &
\colhead{$z_{\rm CO}$} &
\colhead{Distance} &
\colhead{$S_{\rm CO}\Delta v$\tablenotemark{b}} &
\colhead{FWHM\tablenotemark{b}} &
\colhead{R$_{\rm 1/2,CO}$\tablenotemark{c}} &
\colhead{R$_{\rm 1/2,HST}$} &
\colhead{M$_{\rm gas}$\tablenotemark{d}} &
\colhead{M$_{\rm stellar}$\tablenotemark{e}} &
\colhead{$\langle\rm SFR\rangle$\tablenotemark{f}} &
\colhead{$f_{\rm gas}$\tablenotemark{g}} \\
\colhead{} &
\colhead{} &
\colhead{(Mpc)} &
\colhead{(\Jykms)} &
\colhead{(\kms)} &
\colhead{(kpc)} &
\colhead{(kpc)} &
\colhead{($10^{10}$\,\Msol)} &
\colhead{($10^{10}$\,\Msol)} &
\colhead{(\myr)} &
\colhead{}
}
\startdata
J0225--371\tablenotemark{h} & 1.599 &  0.11 & 1.26$\pm$0.10 & 442$\pm$39  & 5.0  & 5.8 & 23.3$\pm$1.8  &  6.3$^{+0.8}_{-0.9}$  & 173$\pm$76 & 0.79$^{+0.02}_{-0.03}$   \\ 
J0225--460\tablenotemark{h} & 1.600 & 0.24 & 0.50$\pm$0.05 & 388$\pm$44 & 2.5  & 1.9 & 9.3$\pm$0.9 &  9.1$^{+6.0}_{-3.5}$  & 116$\pm$60 & 0.50$^{+0.17}_{-0.10}$   \\ 
J0225--281\tablenotemark{h} & 1.611 & 0.23 &  0.80$\pm$0.08 & 292$\pm$34 & 4.0  & 9.0 & 14.8$\pm$1.5 &  5.4$^{+3.5}_{-2.6}$  & 120$\pm$50 & 0.73$^{+0.13}_{-0.10}$   \\ 
J0225--541\tablenotemark{h} & 1.611 &  0.29 & 1.12$\pm$0.26 & 341$\pm$89 & 6.4  & 7.1 & 20.8$\pm$4.7 &  6.6$^{+0.8}_{-0.9}$  &  82$\pm$30 & 0.76$^{+0.05}_{-0.05}$   \\ 
J0225--429 & 1.602 & 0.21 & 0.25$\pm$0.05 & 283$\pm$65 & 3.0  & 1.3 & 4.7$\pm$0.9 &  0.6$^{+1.9}_{-0.1}$  & 178$\pm$83 & 0.89$^{+0.33}_{-0.02}$   \\
J0225--407 & 1.599 & 0.16 & 0.26$\pm$0.04 & 290$\pm$57 & 3.5  & 4.7 & 4.8$\pm$0.8 &  0.7$^{+2.6}_{-0.3}$  & 84$\pm$28 & 0.87$^{+0.41}_{-0.05}$   \\ 
J0225--324 & 1.600 & 0.26 & 0.09$\pm$0.02 & 193$\pm$61 & $<$1.9  & 3.5 & 1.7$\pm$0.4 &  0.1$^{+0.3}_{-0.0}$  & 48$\pm$27 & 0.94$^{+0.14}_{-0.02}$   \\ 
J0225--303\tablenotemark{i} & 1.596 & 0.0 & 0.55$\pm$0.15 & 687$\pm$222 & 4.0  & 5.4 & 10.2$\pm$2.8 &  4.4$^{+0.8}_{-0.9}$  & 3$\pm$3 & 0.70$^{+0.07}_{-0.07}$   \\  
\enddata
\tablenotetext{a}{Distance to the Brightest Cluster Galaxy.}
\tablenotetext{b}{Computed from a Gaussian fit to the spectral profile.}
\tablenotetext{c}{The geometric mean of the semi-major and semi-minor axes (HWHM) from a 2D Gaussian fit.}
\tablenotetext{d}{Calculated using $r_{21} = 0.77$, $\alpha_{\rm CO}=4.36$.}
\tablenotetext{e}{Described in detail in \cite{Nantais16}.}
\tablenotetext{f}{Described in detail in \cite{Noble17}.}
\tablenotetext{g}{Defined as ${\rm M_{gas}/(M_{gas}+M_{stellar})}$.}
\tablenotetext{h}{Sources also presented in  \cite{Noble17}, with values updated slightly based on the Cycle 5 observations and updated photometry.}
\tablenotetext{i}{Merging pair, where the reported CO luminosity, SFR, and effective radii are for a single component.  The stellar mass, measured on the combined system, has been divided by two.  The molecular gas spectral profile is fit with a double Gaussian.}

\end{deluxetable*}

\subsection{CO (2--1) Sizes and Gas Masses}
\label{sec:CO}
The Cycle 5 pointing encompasses four known gas-rich cluster members from \cite{Noble17}, but now detected at higher S$/$N, $\sim10\times$ higher spatial resolution, and finer spectral resolution.  Due to the increased quality of the data, we additionally detect four new sources in CO, all coincident with HST counterparts, and consistent with the cluster redshift for CO (2--1) emission. 

To measure the size of the CO disks we first create S$/$N contours (using the noise measured within the aforementioned annulus) on the integrated-intensity map.  Within a $3\sigma$ region around each source, we model the flux using a two-dimensional Gaussian profile with an elliptical cross-section, obtaining a best-fit major and minor FWHM.  We note that this is roughly equivalent to a half-light radius, as the semi-major and minor FWHM (i.e.\ HWHM) contain 50\% of the flux within a two-dimensional Gaussian profile.  Subtracting the model fit from the integrated-intensity image produces low residuals (all within 0.6 of the local pixel-to-pixel rms)
in each case.  The resulting molecular gas radii, $R_{\rm1/2,CO}$, range from $\sim$2--7\,kpc, defined as the geometric average of the semi-major and semi-minor axis, deconvolved from the beam.  One detection (J0225-324) is only marginally resolved, and we therefore use the clean beam as an upper limit on its size.  The  uncertainties on the half-light gas radii are $\sim5-15\%$, determined from standard propagation of the errors on the best-fit axes.

We adopt the same approach as in \cite{Noble17} to measure the CO flux. Using the aforementioned best-fit major and minor FWHM for each source, we extract spectral profiles defined over a 4$\sigma$ ($\sigma=\rm FWHM/2.355$) region on the full image cube.  We use a Gaussian function to model the CO emission line profile, with the area under the Gaussian corresponding to the integrated flux.  Errors are determined from the rms of the line-free channels, defined as more than $\sim7\sigma$ away from the centroid. From the HST imaging, we have identified one close pair (J0225-303), which is blended in all other ancillary data, and displays clear double velocity components in the CO spectral profile.  We therefore fit a double Gaussian to the data in this case.

From \cite{Solomon05}, we convert the CO flux into a line luminosity using

\begin{equation}
L^{\prime}_{\rm CO} = 3.25\times10^{7}\times S_{\rm CO} \Delta v \frac{D^2_{\rm L}}{\nu_{\rm rest}^{2} (1+z)}\, {\rm K\,km\,s^{-1}\, pc^{2}}
\end{equation}
where $S_{\rm CO} \Delta v$ is the velocity-integrated flux in \Jykms, $D^2_{\rm L}$ is the luminosity distance in Mpc, and $\nu_{\rm rest}$ is the rest-frame frequency of CO (2--1) of 230.54\,GHz.

To estimate the total molecular gas mass, we assume sub-thermalized emission with brightness temperature ratio of 0.77 \citep{Daddi15, Genzel15} between $J=2$ to $J=1$, and an \alphaCO\ conversion factor of 4.36, which is commonly used for normal star-forming galaxies \citep{Solomon91, Bolatto13, Carleton17}:  

\begin{equation}
\frac{M_{\rm gas}}{M_{\sun}}  = \alpha_{\rm CO} \frac{L^{\prime}_{(2-1)}}{0.77}
\end{equation}

Our final sample consists of galaxies with gas masses $\sim 1-25\times10^{10}$\,\Msol.  We note the remeasured fluxes of the four galaxies that were detected in \cite{Noble17} are all within 1.5$\sigma$ of the previous shallower data.  Gas masses, CO sizes, and supporting measurements are presented in Table 1.

\subsection{HST Sizes}
We estimate the size of the optical stellar disk using the highest-resolution near-infrared imaging available, namely the F160W filter on WFC3 aboard \textit{HST}, corresponding to rest-frame $r$-band.  We use a GALFIT \citep{Peng02} wrapper, detailed in Matharu et al.\ (submitted), to fit a single-component S\'ersic profile and measure a half-light radius, $r_{e}$, for each galaxy.  For the case of the galaxy pair (J0225-303), we extract separate half-light radii for each component. 

\subsection{Size Comparisons in the Literature}
\label{sec:lit}
From the literature, we compile CO and optical sizes for galaxies in the coeval field, and additionally in Virgo and the low-$z$ field.  We briefly describe those samples here.

{\it High-z field:} The PHIBSS survey contains six star-forming galaxies with sub-arcsecond resolution CO (3--2) imaging (down to scales of $\sim$5\,kpc) over a redshift slice of $1<z<1.6$ \citep{Tacconi13}.  They additionally extract molecular gas half-light radii for 10 galaxies, and include four $z=1.5$ sources from \cite{Daddi10}, albeit all with larger beams of $\sim$12\,kpc.  The CO sizes constitute a mixture of fits to $n=1$ S\'ersic profiles and circular Gaussians.  We note that fitting a circular Gaussian should be similar to our method of fitting an elliptical Gaussian and circulazing the semi-major and semi-minor axes through the geometric mean.  In all cases, the optical sizes are derived from S\'ersic fits to rest-frame $b$- and $r$-band imaging using GALFIT.  This constitutes our primary field comparision.

{\it Low-z field:}  The BIMA Survey provides 13 galaxies with measured optical disks from $R/J/K/I$ and gas disks from CO (1--0) \citep{Regan01}.  The sizes are tabulated as scale lengths, which we convert into effective half-light radii by multiplying by a factor of 1.678, appropriate for an $n=1$ S\'ersic profile.  We note \cite{Leroy08} also contains molecular gas sizes for many of the same galaxies, but the optical sizes are derived from 3.6\um\ imaging and therefore less applicable for our comparison.

{\it Virgo cluster:}  The largest compilation of CO sizes of Virgo cluster galaxies is from \citet[][but see also \citealt{Pappalardo12, Chung17}]{Kenney88}. In \cite{Kenney88}, they calculate a CO (1--0) effective radius that contains 70\% of the flux, by analytically fitting a model distribution (either a Gaussian or exponential).  We convert this into a half-light radius using the model function and derived scale lengths.  The effective radii for the optical disks of each galaxy are taken from \cite{McDonald09}, using a combined S\'ersic plus exponential fit to the $r$-band surface brightness profile.

\section{Results and Discussion}

\subsection{Rotating Molecular Gas Disks in $z\sim1.6$ Cluster Galaxies}
\label{sec:rot}

\begin{figure*} \centering
\subfigure{\includegraphics[width=\subfigsztwo\columnwidth]{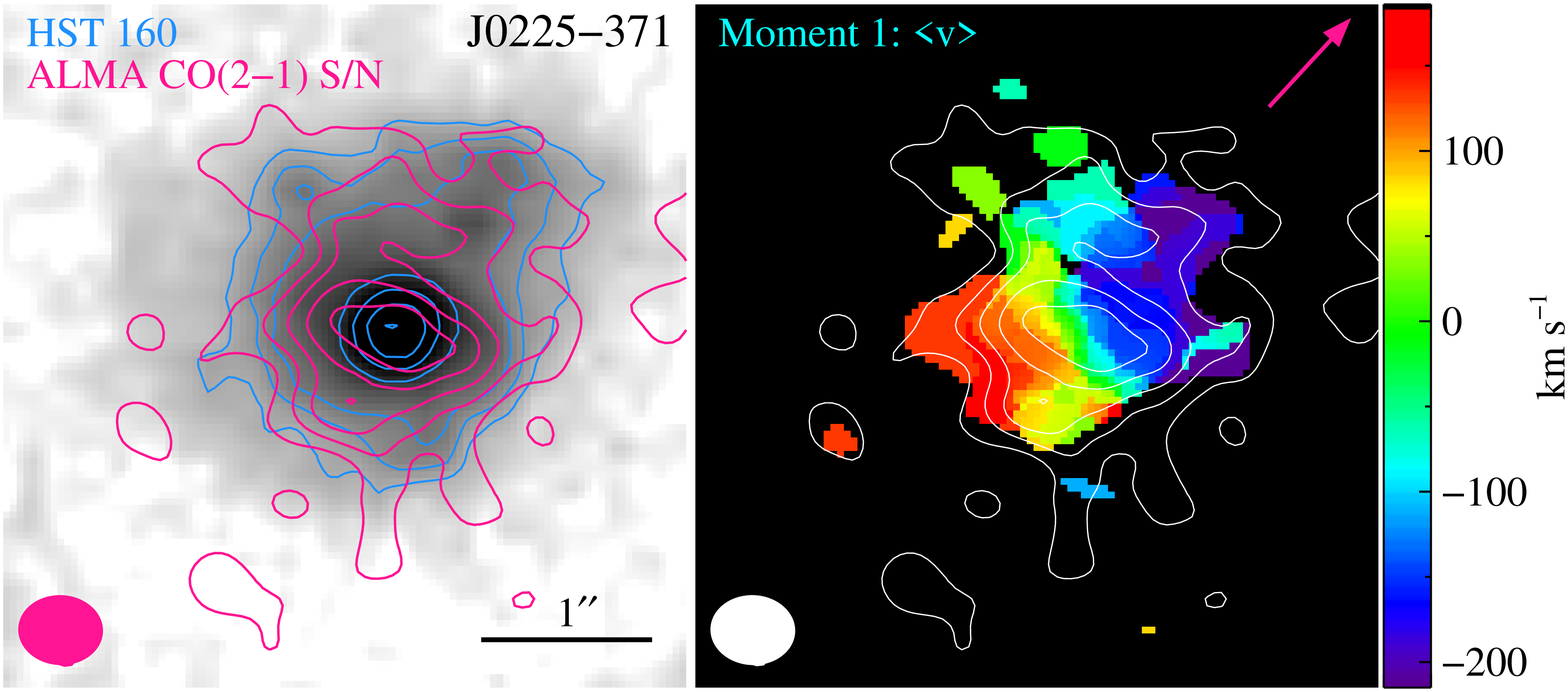}} \hfill
\subfigure{\includegraphics[width=\subfigsztwo\columnwidth]{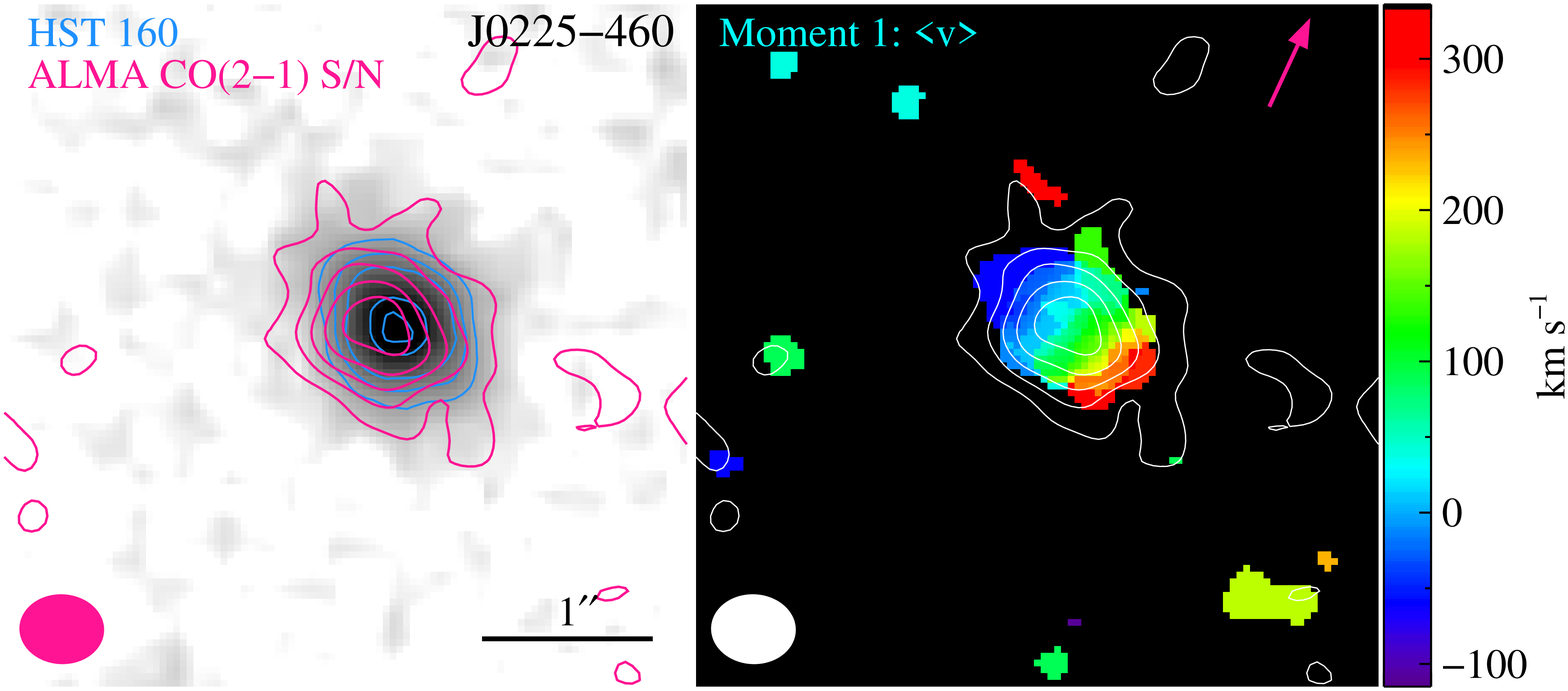}} \hfill
\subfigure{\includegraphics[width=\subfigsztwo\columnwidth]{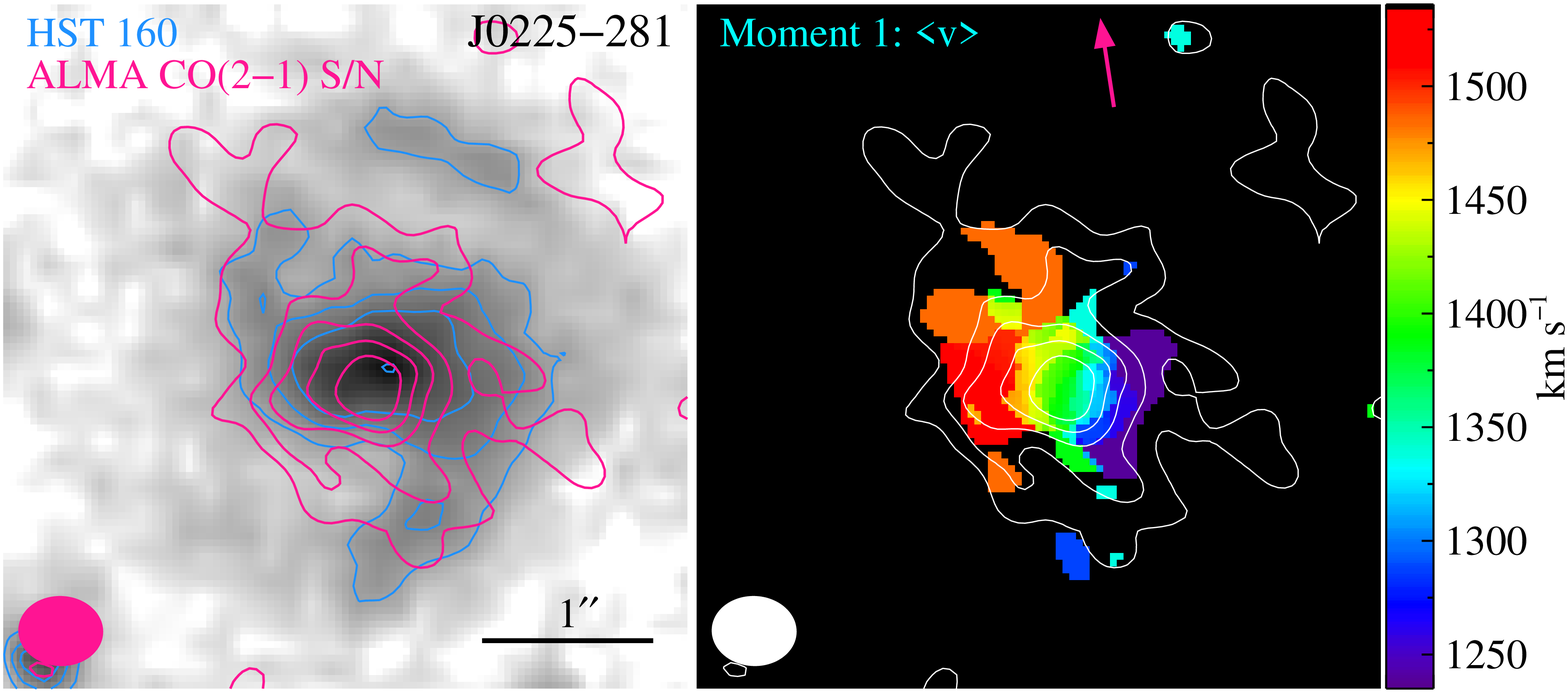}} \hfill
\subfigure{\includegraphics[width=\subfigsztwo\columnwidth]{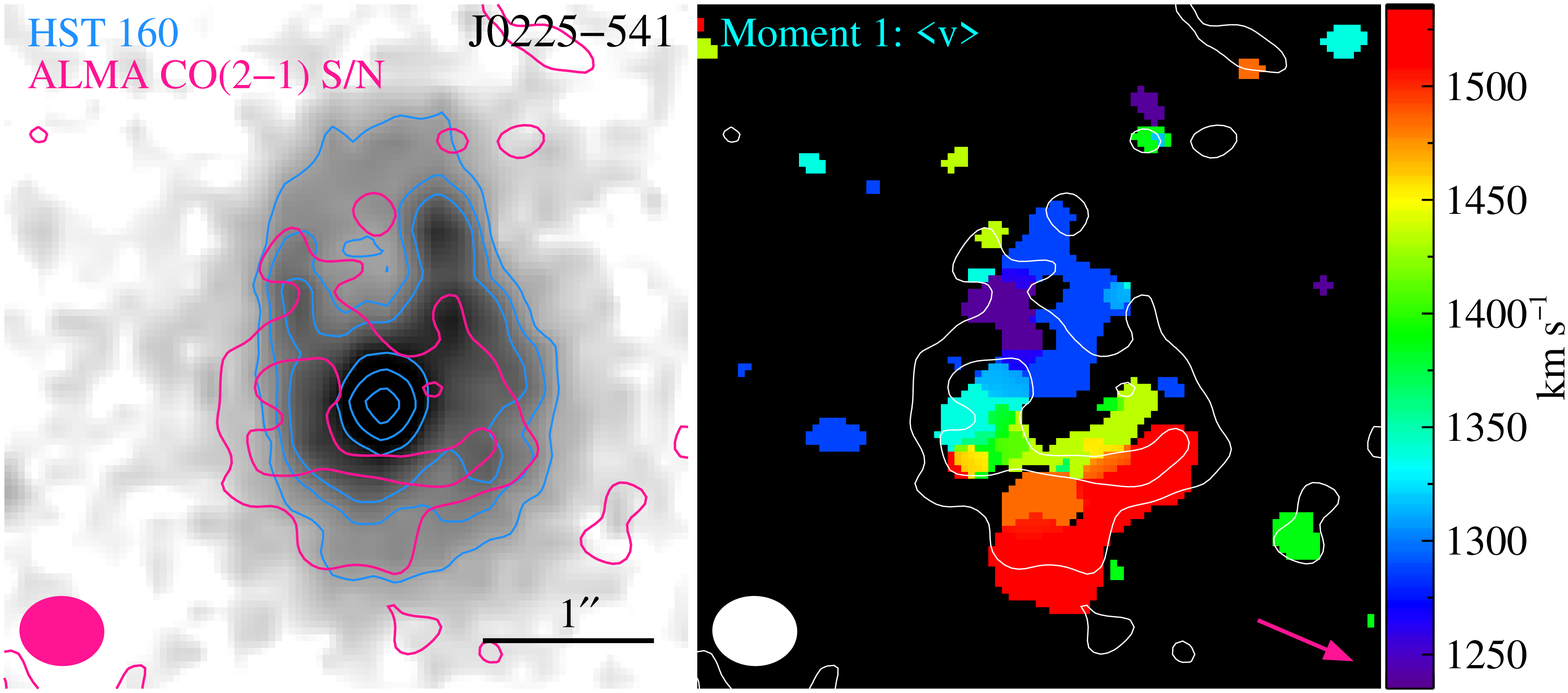}}\hfill
\caption{Integrated-intensity (left panel of each source) and intensity-weighted velocity (right panel of each source) maps for the four highest S$/$N detections, created from 50\,\kms\ channels and spanning 4\asec$\times$4\asec ($\sim34\times34$\,kpc).   The gray-scale image on the left is from HST F160W (highlighted with arbitrary surface brightness contours in blue), with CO integrated-intensity S$/$N contours (in pink) starting at 2$\sigma$ in 2$\sigma$ steps.  The same CO contours are included on the right panel in white, overlaid on the velocity map with the corresponding color bar.  The velocities are relative to the central redshift of source J0225-371, which has the highest S$/$N. The pink arrow in each stamp represents the direction to the cluster center (as determined by the BCG). The synthesized beam of 0.50\asec$\times$0.41\asec ($\sim3.5$\,kpc) is shown as an ellipse in the bottom left corner. In each stamp, north is up and east is to the left.}
 \label{fig:stamps50} 
\end{figure*}

In Figure~\ref{fig:stamps50}, we show integrated-intensity and velocity maps for the four highest S$/$N sources.  In three of the four galaxies, we see indications of rotation in the molecular gas disk, as evidenced by the (mostly) smooth velocity gradients transitioning from blue to red.  In particular, J0225-371, J0225-460, and J0225-281 show a pattern characteristic of ordered rotation (\citealp[e.g.,][]{Sofue01}), with a monotonic velocity gradient and alignment between the kinematic axis and major axis of the stellar disk.  J0225-281 displays slightly more asymmetry, both kinematically and in the outer intensity contours; the {\it HST} image, however, reveals a rather symmetric spiral galaxy.  The most perturbed galaxy is J0225-541, exhibiting a molecular gas void in the center, though a velocity gradient is still present.  A central depletion of CO has also been observed in a single Virgo cluster galaxy, with \cite{Kenney88} attributing it to a molecular gas ring.

\begin{figure*} \centering
\subfigure{\includegraphics[width=\subfigsztwo\columnwidth]{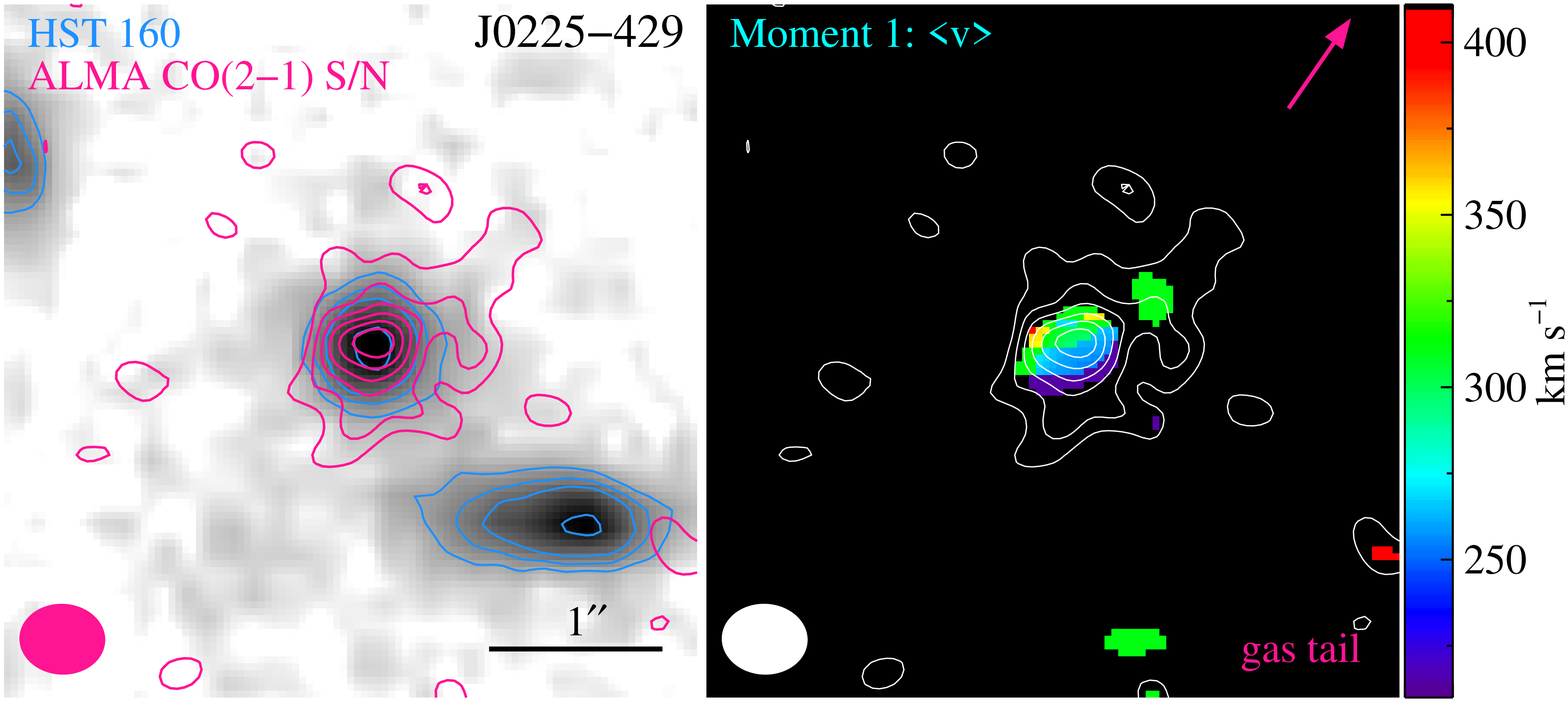}} \hfill
\subfigure{\includegraphics[width=\subfigsztwo\columnwidth]{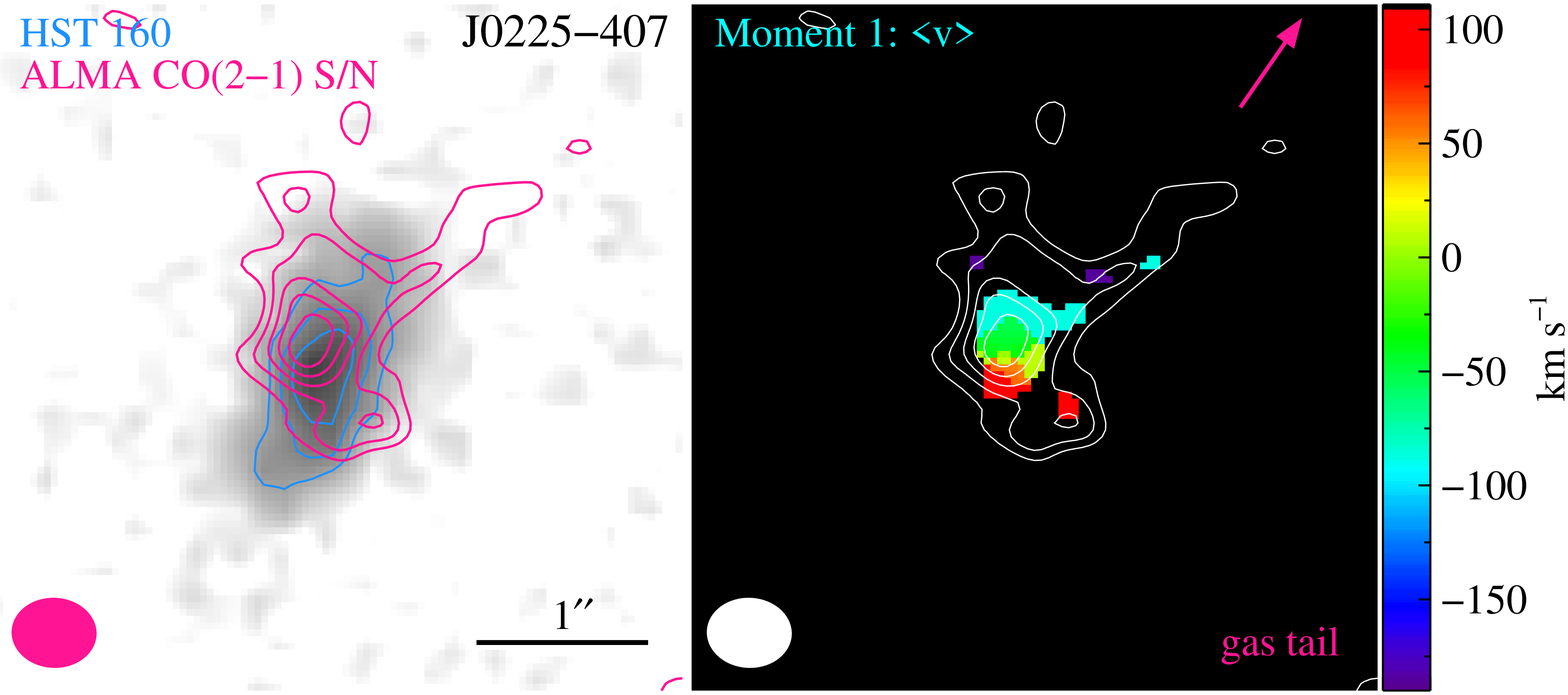}} 
\subfigure{\includegraphics[width=\subfigsztwo\columnwidth]{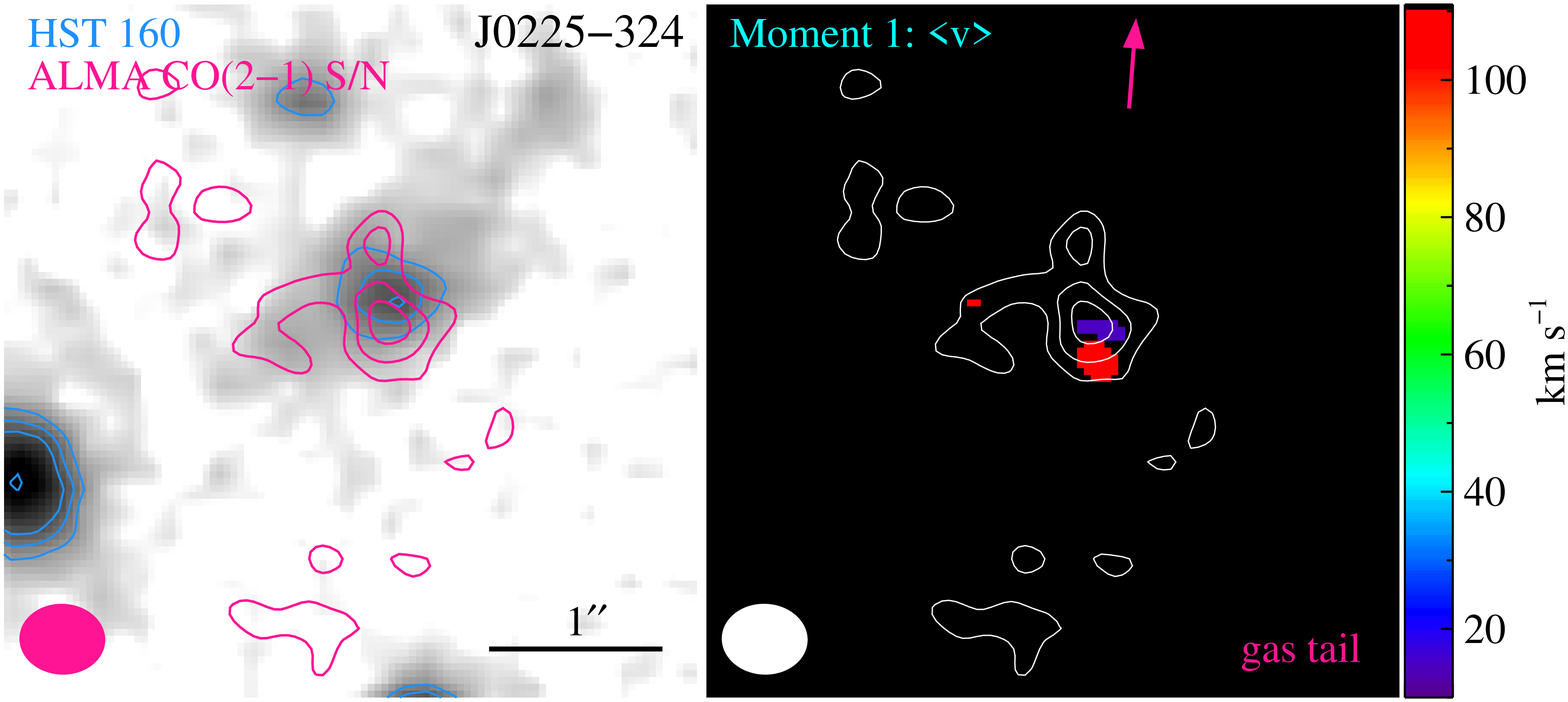}} 
\subfigure{\includegraphics[width=\subfigsztwo\columnwidth]{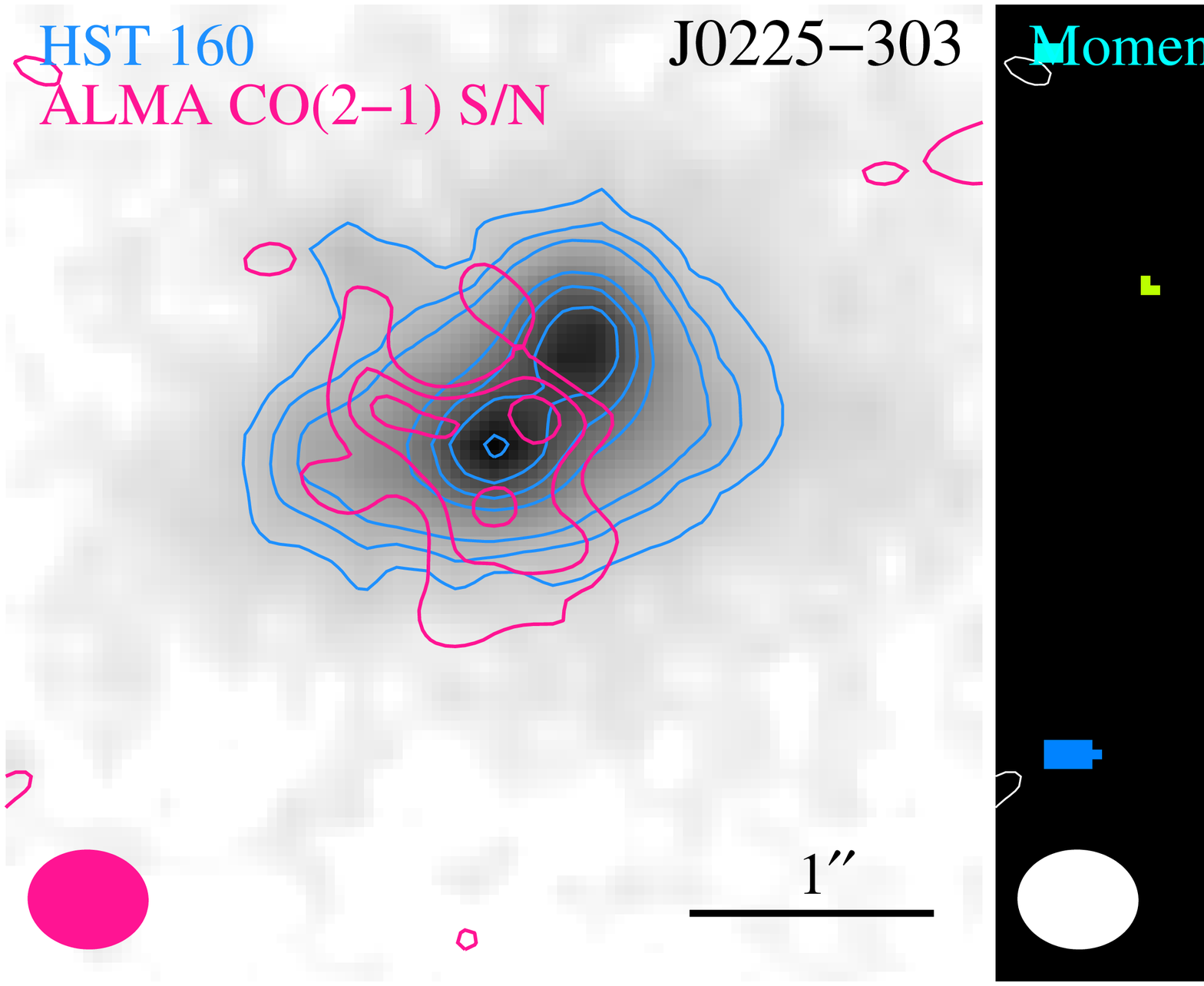}}
\caption{As in Figure~\ref{fig:stamps50}, integrated-intensity (left panel) and intensity-weighted velocity (right panel) maps, but for lower S$/$N sources, created from a velocity resolution of 100\,\kms.  The CO S$/$N contours still start at 2$\sigma$, but now increase in steps of 1$\sigma$.  The velocities are relative to the central redshift of source J0225-371.  The galaxies which display possible one-sided tails are noted. The lower right panel, J0225-303, is the likely BCG.}
 \label{fig:stamps100} 
\end{figure*}

We present the same postage stamps in Figure~\ref{fig:stamps100}, but for the four galaxies detected at lower S$/$N in the ALMA maps.  We again see evidence for a velocity gradient in two galaxies, J0225-429 and J0225-407,  which could be due to velocity shear rather than rotation, though there is no clear evidence of merger events.   The two remaining detections are at too low S$/$N to accurately characterize the velocity maps.

Galaxies J0225-429, J0225-407, and J0225-324 display a one-sided tail in the intensity maps at the 2--3$\sigma$ level.  This is further supported by an elongated kinematic feature in both J0225-407 and J0225-429.  This could be suggestive of ram-pressure stripping, and is reminiscent of gaseous debris tails seen in jellyfish galaxies (e.g.,\ \citealp{Ebeling14, Poggianti16, Sheen17, Moretti18}).  The molecular gas tails roughly point toward the projected cluster center, signifying the galaxies would be stripped on their way back out after reaching pericenter.  Moreover, \cite{Ebeling14} find ram-pressure induced features on galaxies with tangential infall orbits (as opposed to purely radial orbits), resulting in prolonged stripping events. While the {\it HST} images of J0225-429 and 407 look fairly undisturbed, J0225-324 shows some low surface brightness asymmetries, and could be tidally interacting with the faint source to the northeast, which has a photometric redshift of $z\sim1.87$.  We also note there is a $\sim2.5$\,kpc offset in the centroid of the CO emission compared to that of HST in J0225-324 and J0225-407, additionally suggestive of perturbed molecular gas.

The only likely merger in this sample, J0225-303, is also the plausible brightest cluster galaxy (BCG), defined by its total combined $K$-band flux.  The moment 0 and 1 maps show multiple components over the complex, separated spatially and kinematically, over a few arcseconds and $\sim$700\,\kms, respectively (though each component in the double Gaussian fit has a FWHM $\sim$350\,\kms).  As the gas is mostly concentrated over the southern galaxy in the pair, we assume it is primarily emanating from this source.  The massive molecular gas reservoir totals more than $10^{11}$\,\Msol, similar to that of a known starbursting BCG in a $z=1.7$ cluster \citep{Webb17}.  Interestingly, however, the merging system here is mostly devoid of any dusty star formation, suggesting an inefficiency in converting the gas reservoir into stars.  This is consistent with recent findings of post-starburst galaxies that have massive molecular gas reservoirs despite very low star formation rates \citep{French15}.

In summary, we see evidence for velocity gradients in six of the eight cluster galaxies (75\%), albeit some with minor kinematic perturbations (including a molecular gas hole).  At least three of those are likely rotating disks given their smooth, monotonic velocity gradients and alignment of kinematic and structural axes, while the other three display either rotation or velocity shear.  Of the remaining two with no clear velocity gradients, one is a merger and one is only marginally resolved (though has some hints of an asymmetric gas distribution).  This is consistent with the findings of $z\sim1-1.6$ field galaxies from PHIBSS, where two-thirds of the sample are classified as rotating disks of molecular gas.

\subsection{The Extent of Molecular Gas Compared to the Stellar Component}
\label{sec:sizes}

\begin{figure} \centering
\includegraphics[width=1.0\columnwidth]{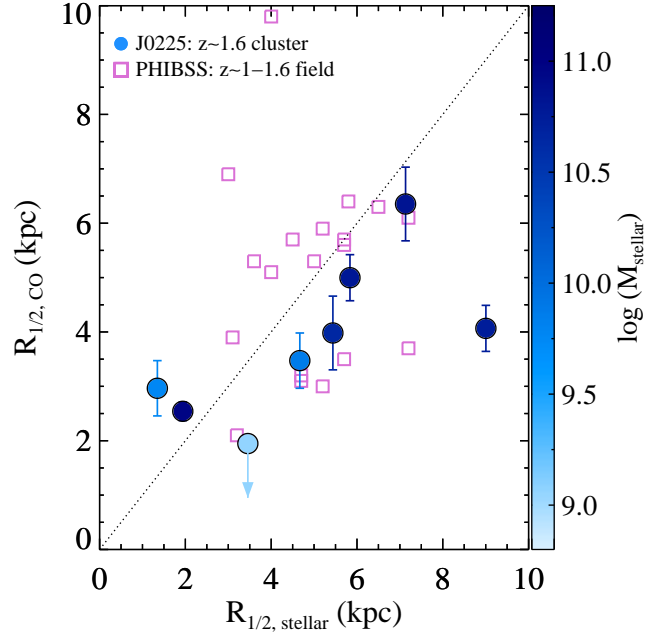}
\caption{The half-light radii of the CO emission compared to the stellar component, measured on the rest-frame optical \textit{HST} image.  The $z\sim1.6$ cluster galaxies in J0225 are shown as circles, color-codes by their stellar mass.  The pink squares show a field comparison sample from PHIBSS \citep{Tacconi13, Daddi10}.  Most of the galaxies fall around the dotted line, representing equal sizes for both components. }
\label{fig:sizes}
\end{figure}

\begin{figure} \centering
\includegraphics[width=0.72\columnwidth]{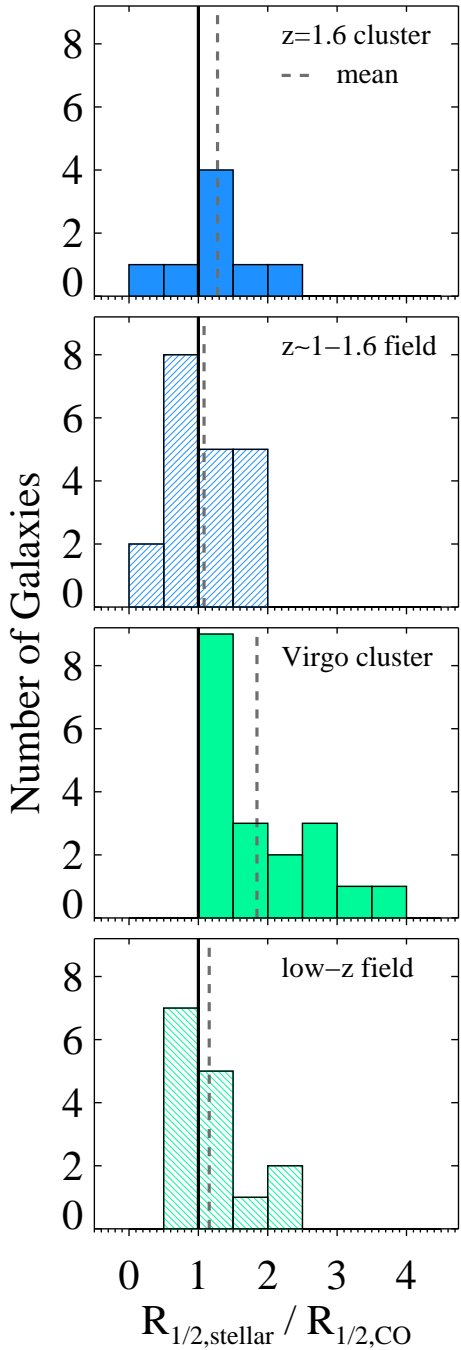}
\caption{The ratio of the half-light radius of the optical stellar component to that of the molecular gas disk. In all panels the solid black line represents a ratio of unity, and the dashed gray line is the mean value for the given population.  The top panel displays our results from J0225 for $z\sim1.6$ cluster galaxies (filled blue histogram).  The blue striped histogram below it is for coeval field galaxies in PHIBSS \citep{Tacconi13, Daddi10}.  The two bottom panels represent lower-$z$ distributions in Virgo cluster galaxies (filled green histogram, \citealp{Kenney88}) and in the field (striped green histogram, \citealp{Regan01}).  There is a tentative trend for cluster galaxies to have smaller gas components compared to the optical stellar component both at low and high redshifts. }
\label{fig:ratio}
\end{figure}

With sub-arcsecond CO observations ($\sim$3.5\,kpc scales at $z\sim1.6$), we can sufficiently compare the size of the molecular gas disk to that of the optical stellar disk, as measured with rest-frame $r$-band from {\it HST} imaging.  In Figure~\ref{fig:sizes}, we show the half-light radii for the $z\sim1.6$ cluster galaxies and a field comparison sample from PHIBSS \citep{Tacconi13}, which also includes sizes for BzK galaxies in \cite{Daddi10}, as described in \S\ref{sec:lit}.  The two samples span a similar range of gas extents, with radii from 2--7\,kpc.  The cluster galaxies mostly lie along a one-to-one ratio of gas-to-stellar sizes, albeit with more galaxies falling below the ratio.

We compute the ratio of the half-light radii of the stellar-to-gas component for the eight $z\sim1.6$ cluster galaxies in the upper panel of Figure~\ref{fig:ratio} (with the black solid line indicating equal sizes).   On average, the cluster CO distributions are slightly smaller than their optical component, with a mean ratio of $1.3\pm0.2$ (using the standard error on the mean), shown by the dashed line.  Moreover, the histogram contains a single upper limit on the CO size for J0225-324, or equivalently, a lower limit on the mean ratio; an actual measurement of its CO size would push the sample mean toward an even higher value.  The two lowest ratios (J0225-460 and 429) are not dominated by extended CO disks, but rather more highly concentrated optical distributions, with S\'ersic indices of 1.9 and 2.7, respectively.  We also note that we have included the merging BCG within this histogram, and have similarly not omitted any mergers in the literature studies described below.  

We compare the cluster distribution to the coeval field PHIBSS sample.  The $1<z<1.6$ field galaxies exhibit roughly equal effective radii between the CO and optical disk components, with a mean ratio of $1.1\pm0.1$.  The field and cluster samples span the same range of gas masses, and thus a comparison between them is primarily limited by small number statistics; we cannot differentiate between the two samples with a Kolmogorov-Smirnov test.  Nevertheless, there is a hint that high-$z$ cluster galaxies have a higher ratio of optical-to-gas size, with a $\sim0.9\sigma$ offset between the mean ratios.

To contextualize any difference in the size ratio between the two high-$z$ environments, we consider larger CO samples in the low-$z$ field \citep{Regan01} and within the local Virgo cluster \citep{Kenney88}.  We note however that these studies probe gas masses roughly two orders of magnitude lower than the high-$z$ samples, reflecting the reduced overall gas fractions at lower redshift.  As in PHIBSS, low-$z$ field galaxies (bottom panel) also exhibit comparable stellar-to-gas sizes ($1.2\pm0.1$), in agreement with other nearby field studies \citep{Young95, Leroy08}.  However, the most striking population is within the Virgo cluster, where the gas radii are systematically smaller than the optical component, including a tail out to a ratio of $\sim4$.  The mean ratio in Virgo ($1.8\pm0.2$) is $\sim3.1\sigma$ offset from the coeval field relation, and the Kolmogorov-Smirnov statistic rejects the null hypothesis at a significance of $>99\%$. While optical and CO radii have been compared before at lower redshifts, it has typically been done as a function of galaxy morphology (\citealp[e.g.,][]{Young95}) or HI-deficiency (\citealp[e.g.,][]{Boselli14, Chung17}), rather than environment.  To our knowledge, we are presenting one of the first quantitative comparisons of the distribution of stellar-to-gas radii in field versus cluster environments, revealing a rather conspicuous difference which might stem from gas stripping in high-density environments.

Indeed, some of the most direct evidence for a marked influence on molecular gas by the ICM stems from spatially-resolved CO studies of Virgo galaxies (\citealp[e.g.,][]{Kenney90, Vollmer08, Lee17}), revealing disturbed CO morphologies, extraplanar molecular gas, clumpy gas kinematics, and compressed CO.  With even larger samples, Virgo galaxies have additionally been shown to have truncated H$\alpha$ disks, further demonstrating that multiple gas components are stripped within cluster environments \citep[e.g.,][]{Koopmann04}. Moreover, dust-enshrouded star-formation is also found to be more centrally concentrated than the stellar disk  \citep{Finn18}.  Many of these results are thus favoring an outside-in quenching mechanism for low-redshift galaxies within dense environments.

As the $z\sim1.6$ cluster environment exhibits the same tendency for galaxies with more compact gas components, this is possibly indicative of molecular gas stripping in cluster environments at both redshifts.  However, the difference in distributions between field and cluster environments is much less pronounced at $z\sim1.6$; this could be suggestive of an immature intracluster medium in the younger cluster environment which has less influence on cluster galaxies.
Resolved molecular gas samples are still in their infancy at higher redshifts, with samples too small to say anything statistically significant.  Moreover, inhomogeneity between molecular gas transitions, resolution scales, optical bands, and imaging depths precludes a direct comparison between the various fields and redshifts.

\subsection{Confirmation of High Gas Fractions}
\label{sec:fgas}

\begin{figure*} \centering
\subfigure{\includegraphics[width=1.0\columnwidth]{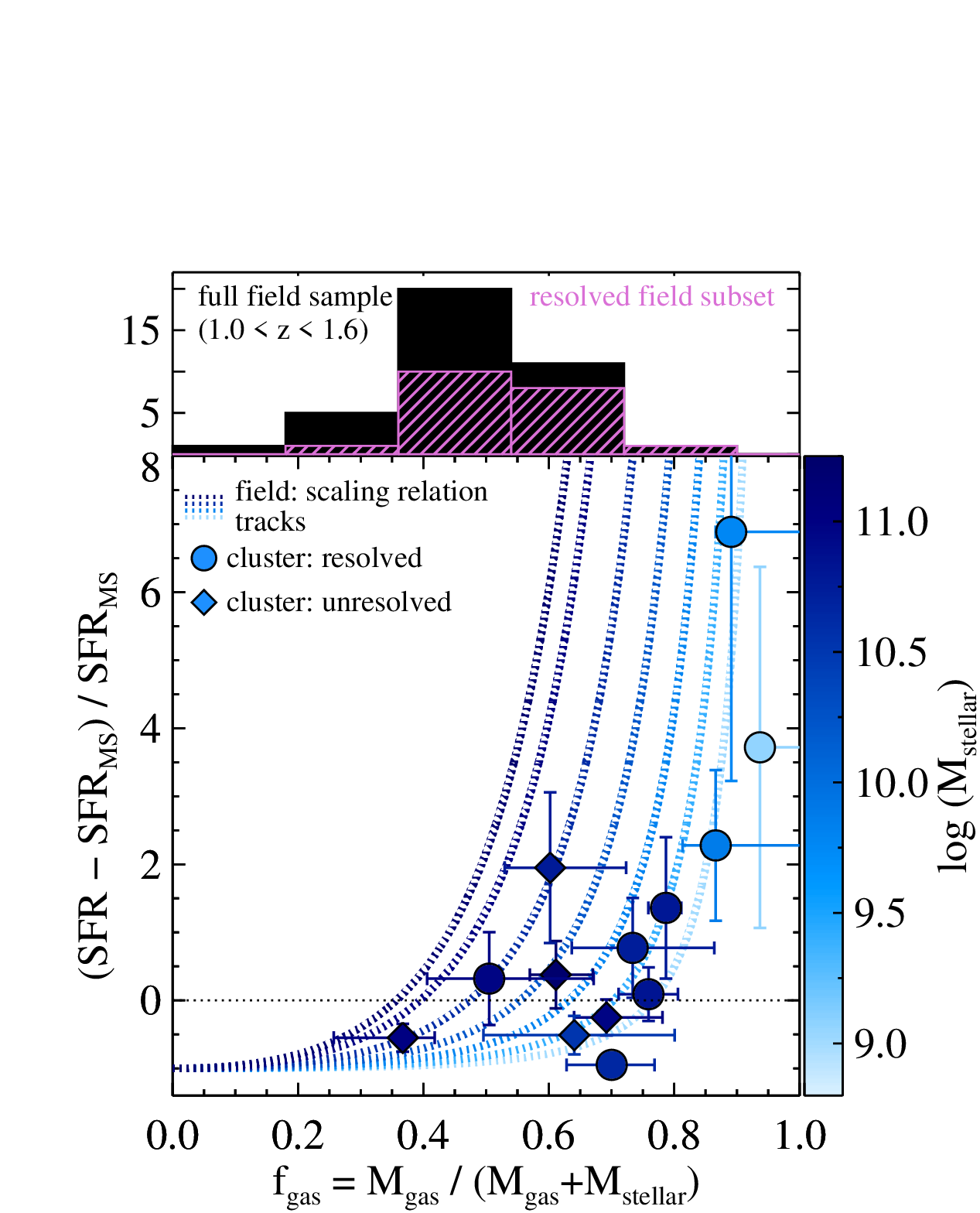}}\hfill
\subfigure{\includegraphics[width=1.0\columnwidth]{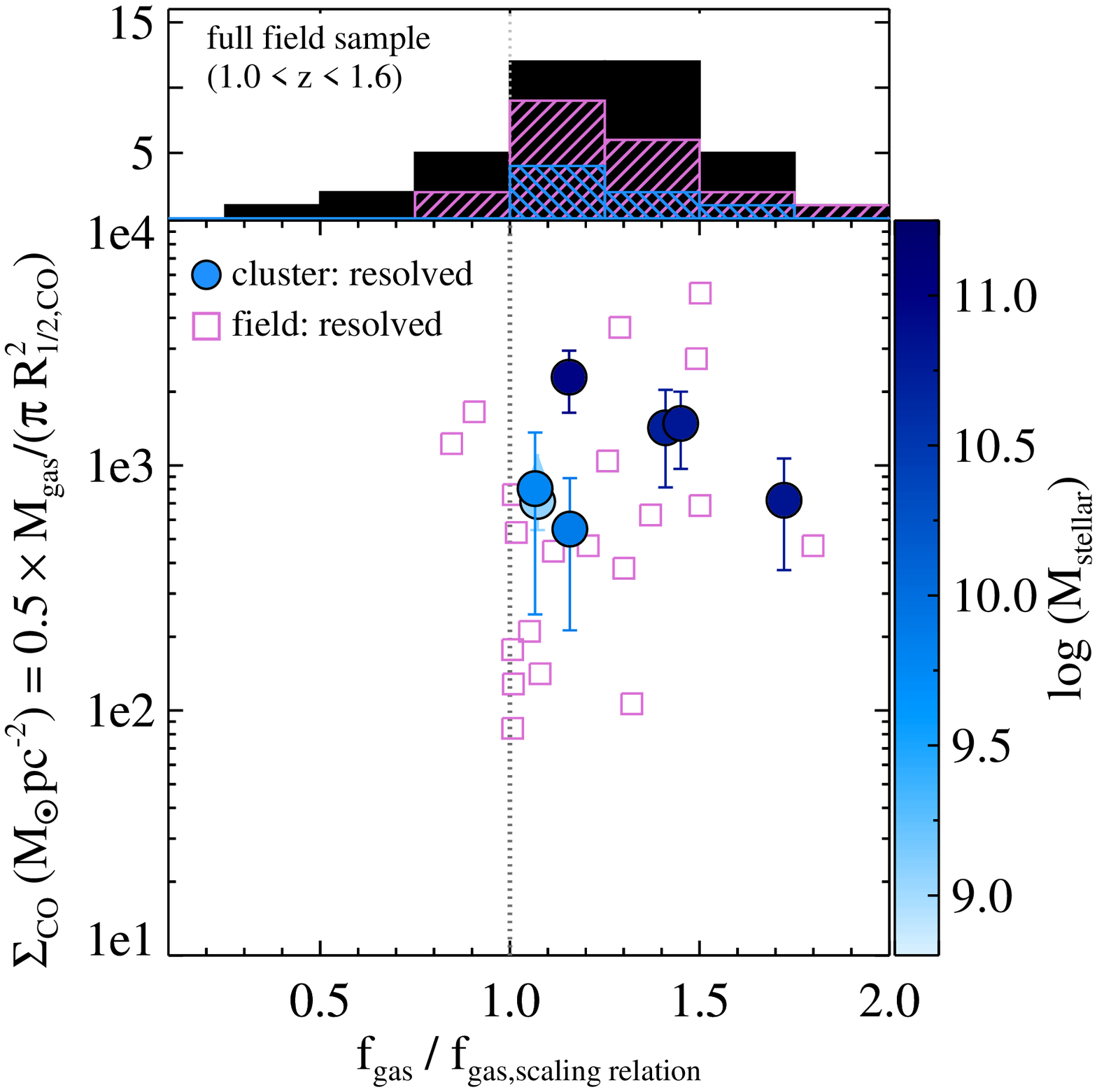}}
\caption{Left---The gas fraction as a function of offset from the star forming main-sequence for the resolved CO detections in this work (filled blue circles), and previous unresolved measurements from \cite{Noble17} (filled diamonds).  The galaxies are color-coded by their stellar mass, shown in the color bar.  The field scaling relation from \cite{Genzel15} is also shown, normalized to $z=1.6$ and at a variety of stellar masses representative of the cluster sample (dashed blue lines).   The upper histogram shows the gas fractions for our field comparison, with black representing the full sample, and pink representing the subset of that with resolved CO observations.  Right---The molecular gas surface densities as a function of gas fraction offset from the field scaling relation, calculated for a given mass, redshift, and star formation rate.  The resolved cluster (blue) and field (pink) samples are systematically above the scaling relation (represented by the gray dotted line).  The upper panel shows the distribution of these subsets compared to the full field sample of resolved and unresolved detections. We note the BCG, J0225-303, lies off the x-axis range on this plot, given its high gas fraction but low star formation rate.   }
\label{fig:fgas}
\end{figure*}

In \cite{Noble17}, we presented tantalizing evidence for numerous gas-rich galaxies with systematically enhanced molecular gas fractions in $z\sim1.6$ clusters compared to coeval field galaxies of similar mass and SFRs (i.e.\ the field scaling relations, \citealp{Genzel15}).  This result has now been corroborated in another cluster at $z=1.46$ \citep{Hayashi18}.  One possible explanation for the apparent elevated gas fractions is that a distinct conversion factor between CO and H$_2$ (known as \alphaCO) may be needed in different environments.  While we have used a value consistent with that of the PHIBSS field sample, a lower \alphaCO\ conversion may be appropriate for increased gas temperatures and velocity dispersions due to, for example, compressed gas and/or mergers (e.g.,\ \citealp{Narayanan12}).  If instead gas fractions are inherently higher in cluster galaxies, another potential explanation could be that environmental pressure increases the formation of molecular gas through compression of the interstellar medium (e.g.,\ \citealp{Bahe12}). 

In Figure~\ref{fig:fgas}, we again confirm the existence of higher gas fractions in $z\sim1.6$ cluster galaxies, but now with more data and at a higher significance of $\sim 7\sigma$ (left panel).  To shed light on this gas fraction offset, we ultimately want to compare the molecular gas sizes and densities of these cluster galaxies with high gas fractions to coeval field galaxies on the scaling relation; however, the currently available subset of PHIBSS field galaxies with sub-arcsecond CO observations \citep{Tacconi13} is also biased toward higher gas fractions (see right panel).  While the resolved field (pink) and cluster (blue) galaxies roughly span the same range of molecular gas densities, they are all (barring one) systematically biased above the typical field gas fraction, as measured by the scaling relation in \cite{Genzel15}.  This is perhaps not surprising, as the most gas-rich galaxies might have been prioritized for the higher-resolution data in the field. This is even more evident in the histogram in the upper right panel, which compares the full $1<z<1.6$ field sample distribution of gas fractions (relative to the scaling relation) to that of resolved field detections. The subset of resolved field galaxies is lacking a representative sample of galaxies with lower gas fractions.

We therefore are not in a position to definitively determine the source of systematically high gas fractions in clusters for now, as a more representative field sample with sub-arcsecond observations is still required.  We note that additional high-resolution imaging of the star-forming component, in order to expose where the gas is situated within these galaxies and how fast it is being consumed, would also certainly help elucidate this trend in cluster gas fractions.

\subsection{Interpretation}
\label{sec:int}
In summary, at first glance, the $z\sim1.6$ cluster galaxies generally look like galaxies infalling from the field, with typical main-sequence star formation rates and massive molecular gas reservoirs situated in rotating disks.  However, there are potentially important differences from their field counterparts, including elevated gas fractions, slightly smaller CO disks, and possible gas tails, which might all be caused by environmental processes.  

Though speculative, the results presented above, in tandem, might indicate that $z\sim1.6$ cluster galaxies are undergoing modest molecular gas stripping. This could explain the apparently high gas fractions: if the stripping causes the gas to become compressed at the leading edge, a lower \alphaCO\ conversion is required, which would in turn lower the inherent mass of the gas reservoir.  Indeed, \cite{Moretti18} found large amounts of molecular gas in the disks of $z\sim0.05$  jellyfish galaxies, along with CO gas in the tails that likely formed \textit{in situ}. Thus, high gas fractions could result as the stripped gas condenses to form new stars.  Moreover, an environmental influence on CO has been reported in $z<0.02$ cluster galaxies \citep[e.g.,][]{Fumagalli09, Scott13, Kenney90, Vollmer08, Lee17, Mok17}, and extending out to intermediate-redshift clusters at $z\sim0.5$ 
(\citealt{Jablonka13}, but see also \citealt{Stark86, KenneyYoung89, Casoli91} for no differences in CO between field and cluster environments).  At $z\sim1.6$ we might be witnessing the initial stages of environmental effects on molecular gas in early cluster formation.

\section{Conclusions}
We present the first high-resolution, kinematic study of molecular gas in high-redshift clusters to date.  With sub-arcsecond imaging of CO (2--1) on $\sim$3.5\,kpc scales and 50--100\,\kms\ velocity resolution, we have characterized the fraction of rotating disks and measured the extent of molecular gas reservoirs in a sample of eight cluster galaxies at $z\sim1.6$.  Our main results are summarized as follows:

\begin{enumerate}

\item{The majority of the galaxies in our sample ($\sim75\%$) display evidence for velocity gradients, either due to ordered disk rotation or velocity shear.  We also detect molecular gas in the BCG, a merger with multiple velocity components, and one galaxy with an asymmetric CO distribution around a central gas void.  A few galaxies display hints of molecular gas tails and gas-to-stellar centroid offsets, reminiscent of stripping. }

\item{On average, $z\sim1.6$ cluster galaxies have slightly smaller CO extents compared to their optical stellar component, as measured by their half-light radii.  This is more pronounced than in the PHIBSS field comparison sample, with an stellar-to-gas ratio offset of $\sim0.9\sigma$ between the two samples.  As this trend is also consistent with that at lower-redshift, comparing Virgo cluster galaxies to the nearby field, it could be indicative of gas stripping within cluster environments, especially in conjunction with the possible gas tails.  However, direct comparisons between the environments are riddled with caveats due to inhomogeneous data sets and small sample sizes.}

\item{We confirm the existence of systematically higher gas fractions in $z\sim1.6$ cluster galaxies compared to the field scaling relations, as shown in \cite{Noble17} and corroborated in \cite{Hayashi18}.  Due to a lack of representative field samples, the origin of the excess is still uncertain.}

\end{enumerate}

The study of distant molecular gas is a burgeoning field; with the confluence of galaxy clusters and a large ALMA primary beam, we can now reap the benefits of multiplexing to spatially resolve this molecular gas in large samples of high-redshift cluster galaxies for the first time.   Kinematically, the majority of the $z\sim1.6$ cluster galaxies resemble rotating gas disks, similar to field galaxies.  Therefore, in the absence of any environmental influences, high-redshift galaxy clusters, with an abundance of massive molecular gas reservoirs in close proximity to one another, offer the most efficient means of studying kinematic and structural properties of CO.  With this first look, we have measured the size of the gas component; however robust evidence for any environmental trends will require larger samples.  Observations of spatially-resolved CO in additional $z\sim1.6$ cluster galaxies are forthcoming in an approved ALMA Cycle 6 program; with a statistical sample, we will perform a complete kinematic analysis to fully characterize rotation, dynamical masses, and gas surface densities in high-redshift cluster galaxies.

\acknowledgments

This work is supported by the National Science Foundation through grant AST-1517863, by HST program numbers GO-13677/14327.01 and GO-15294, and by grant number 80NSSC17K0019 issued through the NASA Astrophysics Data Analysis Program (ADAP). Support for program numbers GO- 13677/14327.01 and GO-15294 was provided by NASA through a grant from the Space Telescope Science Institute, which is operated by the Association of Universities for Research in Astronomy, Incorporated, under NASA contract NAS5-26555.  R.D. gratefully acknowledges support from the Chilean Centro de Excelencia en Astrof\'isica y Tecnolog\'ias Afines (CATA) BASAL grant AFB-170002. J.N. is supported by Universidad Andres Bello internal research grant no. DI-18-17/RG.

This paper makes use of the following ALMA data: ADS/JAO.ALMA\#2017.1.01228.S and \\
ADS/JAO.ALMA\#2015.1.01151.S. ALMA is a partnership of ESO (representing its member states), NSF (USA) and NINS (Japan), together with NRC (Canada) and NSC and ASIAA (Taiwan) and KASI (Republic of Korea), in cooperation with the Republic of Chile. The Joint ALMA Observatory is operated by ESO, AUI/NRAO and NAOJ.

\bibliography{references}

\begin{thebibliography}{81}
\expandafter\ifx\csname natexlab\endcsname\relax\def\natexlab#1{#1}\fi

\bibitem[{{Bah{\'e}} {et~al.}(2012){Bah{\'e}}, {McCarthy}, {Crain}, \&
  {Theuns}}]{Bahe12}
{Bah{\'e}}, Y.~M., {McCarthy}, I.~G., {Crain}, R.~A., \& {Theuns}, T. 2012,
  \mnras, 424, 1179

\bibitem[{{Bolatto} {et~al.}(2013){Bolatto}, {Wolfire}, \& {Leroy}}]{Bolatto13}
{Bolatto}, A.~D., {Wolfire}, M., \& {Leroy}, A.~K. 2013, \araa, 51, 207

\bibitem[{{Boselli} {et~al.}(2014){Boselli}, {Cortese}, {Boquien}, {Boissier},
  {Catinella}, {Gavazzi}, {Lagos}, \& {Saintonge}}]{Boselli14}
{Boselli}, A., {Cortese}, L., {Boquien}, M., {Boissier}, S., {Catinella}, B.,
  {Gavazzi}, G., {Lagos}, C., \& {Saintonge}, A. 2014, \aap, 564, A67

\bibitem[{{Bothwell} {et~al.}(2010){Bothwell}, {Chapman}, {Tacconi}, {Smail},
  {Ivison}, {Casey}, {Bertoldi}, {Beswick}, {Biggs}, {Blain}, {Cox}, {Genzel},
  {Greve}, {Kennicutt}, {Muxlow}, {Neri}, \& {Omont}}]{Bothwell10}
{Bothwell}, M.~S., {et~al.} 2010, \mnras, 405, 219

\bibitem[{{Brodwin} {et~al.}(2013){Brodwin}, {Stanford}, {Gonzalez}, {Zeimann},
  {Snyder}, {Mancone}, {Pope}, {Eisenhardt}, {Stern}, {Alberts}, {Ashby},
  {Brown}, {Chary}, {Dey}, {Galametz}, {Gettings}, {Jannuzi}, {Miller},
  {Moustakas}, \& {Moustakas}}]{Brodwin13}
{Brodwin}, M., {et~al.} 2013, \apj, 779, 138

\bibitem[{{Bundy} {et~al.}(2015){Bundy}, {Bershady}, {Law}, {Yan}, {Drory},
  {MacDonald}, {Wake}, {Cherinka}, {S{\'a}nchez-Gallego}, {Weijmans}, {Thomas},
  {Tremonti}, {Masters}, {Coccato}, {Diamond-Stanic}, {Arag{\'o}n-Salamanca},
  {Avila-Reese}, {Badenes}, {Falc{\'o}n-Barroso}, {Belfiore}, {Bizyaev},
  {Blanc}, {Bland-Hawthorn}, {Blanton}, {Brownstein}, {Byler}, {Cappellari},
  {Conroy}, {Dutton}, {Emsellem}, {Etherington}, {Frinchaboy}, {Fu}, {Gunn},
  {Harding}, {Johnston}, {Kauffmann}, {Kinemuchi}, {Klaene}, {Knapen},
  {Leauthaud}, {Li}, {Lin}, {Maiolino}, {Malanushenko}, {Malanushenko}, {Mao},
  {Maraston}, {McDermid}, {Merrifield}, {Nichol}, {Oravetz}, {Pan}, {Parejko},
  {Sanchez}, {Schlegel}, {Simmons}, {Steele}, {Steinmetz}, {Thanjavur},
  {Thompson}, {Tinker}, {van den Bosch}, {Westfall}, {Wilkinson}, {Wright},
  {Xiao}, \& {Zhang}}]{Bundy15}
{Bundy}, K., {et~al.} 2015, \apj, 798, 7

\bibitem[{{Carleton} {et~al.}(2017){Carleton}, {Cooper}, {Bolatto}, {Bournaud},
  {Combes}, {Freundlich}, {Garcia-Burillo}, {Genzel}, {Neri}, {Tacconi},
  {Sandstrom}, {Weiner}, \& {Weiss}}]{Carleton17}
{Carleton}, T., {et~al.} 2017, \mnras, 467, 4886

\bibitem[{{Casoli} {et~al.}(1991){Casoli}, {Boisse}, {Combes}, \&
  {Dupraz}}]{Casoli91}
{Casoli}, F., {Boisse}, P., {Combes}, F., \& {Dupraz}, C. 1991, \aap, 249, 359

\bibitem[{{Castignani} {et~al.}(2018){Castignani}, {Combes}, {Salom{\'e}},
  {Andreon}, {Pannella}, {Heywood}, {Trinchieri}, {Cicone}, {Davies}, {Owen},
  \& {Raichoor}}]{Castignani18}
{Castignani}, G., {et~al.} 2018, ArXiv e-prints

\bibitem[{{Chabrier}(2003)}]{Chabrier03}
{Chabrier}, G. 2003, \pasp, 115, 763

\bibitem[{{Chung} {et~al.}(2017){Chung}, {Yun}, {Verheijen}, \&
  {Chung}}]{Chung17}
{Chung}, E.~J., {Yun}, M.~S., {Verheijen}, M.~A.~W., \& {Chung}, A. 2017, \apj,
  843, 50

\bibitem[{{Cibinel} {et~al.}(2017){Cibinel}, {Daddi}, {Bournaud}, {Sargent},
  {le Floc'h}, {Magdis}, {Pannella}, {Rujopakarn}, {Juneau}, {Zanella}, {Duc},
  {Oesch}, {Elbaz}, {Jagannathan}, {Nyland}, \& {Wang}}]{Cibinel17}
{Cibinel}, A., {et~al.} 2017, \mnras, 469, 4683

\bibitem[{{Coogan} {et~al.}(2018){Coogan}, {Daddi}, {Sargent}, {Strazzullo},
  {Valentino}, {Gobat}, {Magdis}, {Bethermin}, {Pannella}, {Onodera}, {Liu},
  {Cimatti}, {Dannerbauer}, {Carollo}, {Renzini}, \& {Tremou}}]{Coogan18}
{Coogan}, R.~T., {et~al.} 2018, \mnras, 479, 703

\bibitem[{{Croom} {et~al.}(2012){Croom}, {Lawrence}, {Bland-Hawthorn},
  {Bryant}, {Fogarty}, {Richards}, {Goodwin}, {Farrell}, {Miziarski}, {Heald},
  {Jones}, {Lee}, {Colless}, {Brough}, {Hopkins}, {Bauer}, {Birchall}, {Ellis},
  {Horton}, {Leon-Saval}, {Lewis}, {L{\'o}pez-S{\'a}nchez}, {Min}, {Trinh}, \&
  {Trowland}}]{Croom12}
{Croom}, S.~M., {et~al.} 2012, \mnras, 421, 872

\bibitem[{{Daddi} {et~al.}(2010){Daddi}, {Bournaud}, {Walter}, {Dannerbauer},
  {Carilli}, {Dickinson}, {Elbaz}, {Morrison}, {Riechers}, {Onodera}, {Salmi},
  {Krips}, \& {Stern}}]{Daddi10}
{Daddi}, E., {et~al.} 2010, \apj, 713, 686

\bibitem[{{Daddi} {et~al.}(2015){Daddi}, {Dannerbauer}, {Liu}, {Aravena},
  {Bournaud}, {Walter}, {Riechers}, {Magdis}, {Sargent}, {B{\'e}thermin},
  {Carilli}, {Cibinel}, {Dickinson}, {Elbaz}, {Gao}, {Gobat}, {Hodge}, \&
  {Krips}}]{Daddi15}
---. 2015, \aap, 577, A46

\bibitem[{{Dekel} {et~al.}(2009){Dekel}, {Birnboim}, {Engel}, {Freundlich},
  {Goerdt}, {Mumcuoglu}, {Neistein}, {Pichon}, {Teyssier}, \&
  {Zinger}}]{Dekel09}
{Dekel}, A., {et~al.} 2009, \nat, 457, 451

\bibitem[{{Demarco} {et~al.}(2010){Demarco}, {Wilson}, {Muzzin}, {Lacy},
  {Surace}, {Yee}, {Hoekstra}, {Blindert}, \& {Gilbank}}]{Demarco10}
{Demarco}, R., {et~al.} 2010, \apj, 711, 1185

\bibitem[{{Ebeling} {et~al.}(2014){Ebeling}, {Stephenson}, \&
  {Edge}}]{Ebeling14}
{Ebeling}, H., {Stephenson}, L.~N., \& {Edge}, A.~C. 2014, \apjl, 781, L40

\bibitem[{{Finn} {et~al.}(2018){Finn}, {Desai}, {Rudnick}, {Balogh}, {Haynes},
  {Jablonka}, {Koopmann}, {Moustakas}, {Peng}, {Poggianti}, {Rines}, \&
  {Zaritsky}}]{Finn18}
{Finn}, R.~A., {et~al.} 2018, \apj, 862, 149

\bibitem[{{F{\"o}rster Schreiber} {et~al.}(2006){F{\"o}rster Schreiber},
  {Genzel}, {Lehnert}, {Bouch{\'e}}, {Verma}, {Erb}, {Shapley}, {Steidel},
  {Davies}, {Lutz}, {Nesvadba}, {Tacconi}, {Eisenhauer}, {Abuter}, {Gilbert},
  {Gillessen}, \& {Sternberg}}]{Forster06}
{F{\"o}rster Schreiber}, N.~M., {et~al.} 2006, \apj, 645, 1062

\bibitem[{{F{\"o}rster Schreiber} {et~al.}(2009){F{\"o}rster Schreiber},
  {Genzel}, {Bouch{\'e}}, {Cresci}, {Davies}, {Buschkamp}, {Shapiro},
  {Tacconi}, {Hicks}, {Genel}, {Shapley}, {Erb}, {Steidel}, {Lutz},
  {Eisenhauer}, {Gillessen}, {Sternberg}, {Renzini}, {Cimatti}, {Daddi},
  {Kurk}, {Lilly}, {Kong}, {Lehnert}, {Nesvadba}, {Verma}, {McCracken},
  {Arimoto}, {Mignoli}, \& {Onodera}}]{Forster09}
---. 2009, \apj, 706, 1364

\bibitem[{{F{\"o}rster Schreiber} {et~al.}(2018){F{\"o}rster Schreiber},
  {Renzini}, {Mancini}, {Genzel}, {Bouch{\'e}}, {Cresci}, {Hicks}, {Lilly},
  {Peng}, {Burkert}, {Carollo}, {Cimatti}, {Daddi}, {Davies}, {Genel}, {Kurk},
  {Lang}, {Lutz}, {Mainieri}, {McCracken}, {Mignoli}, {Naab}, {Oesch},
  {Pozzetti}, {Scodeggio}, {Shapiro Griffin}, {Shapley}, {Sternberg},
  {Tacchella}, {Tacconi}, {Wuyts}, \& {Zamorani}}]{Forster18}
---. 2018, ArXiv e-prints

\bibitem[{{French} {et~al.}(2015){French}, {Yang}, {Zabludoff}, {Narayanan},
  {Shirley}, {Walter}, {Smith}, \& {Tremonti}}]{French15}
{French}, K.~D., {Yang}, Y., {Zabludoff}, A., {Narayanan}, D., {Shirley}, Y.,
  {Walter}, F., {Smith}, J.-D., \& {Tremonti}, C.~A. 2015, \apj, 801, 1

\bibitem[{{Fumagalli} {et~al.}(2009){Fumagalli}, {Krumholz}, {Prochaska},
  {Gavazzi}, \& {Boselli}}]{Fumagalli09}
{Fumagalli}, M., {Krumholz}, M.~R., {Prochaska}, J.~X., {Gavazzi}, G., \&
  {Boselli}, A. 2009, \apj, 697, 1811

\bibitem[{{Genzel} {et~al.}(2003){Genzel}, {Baker}, {Tacconi}, {Lutz}, {Cox},
  {Guilloteau}, \& {Omont}}]{Genzel03}
{Genzel}, R., {Baker}, A.~J., {Tacconi}, L.~J., {Lutz}, D., {Cox}, P.,
  {Guilloteau}, S., \& {Omont}, A. 2003, \apj, 584, 633

\bibitem[{{Genzel} {et~al.}(2010){Genzel}, {Tacconi}, {Gracia-Carpio},
  {Sternberg}, {Cooper}, {Shapiro}, {Bolatto}, {Bouch{\'e}}, {Bournaud},
  {Burkert}, {Combes}, {Comerford}, {Cox}, {Davis}, {Schreiber},
  {Garcia-Burillo}, {Lutz}, {Naab}, {Neri}, {Omont}, {Shapley}, \&
  {Weiner}}]{Genzel10}
{Genzel}, R., {et~al.} 2010, \mnras, 407, 2091

\bibitem[{{Genzel} {et~al.}(2013){Genzel}, {Tacconi}, {Kurk}, {Wuyts},
  {Combes}, {Freundlich}, {Bolatto}, {Cooper}, {Neri}, {Nordon}, {Bournaud},
  {Burkert}, {Comerford}, {Cox}, {Davis}, {F{\"o}rster Schreiber},
  {Garc{\'{\i}}a-Burillo}, {Gracia-Carpio}, {Lutz}, {Naab}, {Newman},
  {Saintonge}, {Shapiro Griffin}, {Shapley}, {Sternberg}, \&
  {Weiner}}]{Genzel13}
---. 2013, \apj, 773, 68

\bibitem[{{Genzel} {et~al.}(2015){Genzel}, {Tacconi}, {Lutz}, {Saintonge},
  {Berta}, {Magnelli}, {Combes}, {Garc{\'{\i}}a-Burillo}, {Neri}, {Bolatto},
  {Contini}, {Lilly}, {Boissier}, {Boone}, {Bouch{\'e}}, {Bournaud}, {Burkert},
  {Carollo}, {Colina}, {Cooper}, {Cox}, {Feruglio}, {F{\"o}rster Schreiber},
  {Freundlich}, {Gracia-Carpio}, {Juneau}, {Kovac}, {Lippa}, {Naab}, {Salome},
  {Renzini}, {Sternberg}, {Walter}, {Weiner}, {Weiss}, \& {Wuyts}}]{Genzel15}
---. 2015, \apj, 800, 20

\bibitem[{{Gonz{\'a}lez-L{\'o}pez} {et~al.}(2017){Gonz{\'a}lez-L{\'o}pez},
  {Barrientos}, {Gladders}, {Wuyts}, {Rigby}, {Sharon}, {Aravena}, {Bayliss},
  \& {Ibar}}]{Gonzalez17}
{Gonz{\'a}lez-L{\'o}pez}, J., {et~al.} 2017, \apjl, 846, L22

\bibitem[{{Hayashi} {et~al.}(2017){Hayashi}, {Kodama}, {Kohno}, {Yamaguchi},
  {Tadaki}, {Hatsukade}, {Koyama}, {Shimakawa}, {Tamura}, \&
  {Suzuki}}]{Hayashi17}
{Hayashi}, M., {et~al.} 2017, \apjl, 841, L21

\bibitem[{{Hayashi} {et~al.}(2018){Hayashi}, {Tadaki}, {Kodama}, {Kohno},
  {Yamaguchi}, {Hatsukade}, {Koyama}, {Shimakawa}, {Tamura}, \&
  {Suzuki}}]{Hayashi18}
---. 2018, \apj, 856, 118

\bibitem[{{Herrera-Camus} {et~al.}(2018){Herrera-Camus}, {Tacconi}, {Genzel},
  {Foerster Schreiber}, {Lutz}, {Bolatto}, {Wuyts}, {Renzini}, {Lilly},
  {Belli}, {Uebler}, {Shimizu}, {Davies}, {Sturm}, {Combes}, {Freundlich},
  {Garcia-Burillo}, {Cox}, {Burkert}, {Naab}, {Colina}, {Saintonge}, {Cooper},
  {Feruglio}, \& {Weiss}}]{Herrera18}
{Herrera-Camus}, R., {et~al.} 2018, ArXiv e-prints

\bibitem[{{Hodge} {et~al.}(2012){Hodge}, {Carilli}, {Walter}, {de Blok},
  {Riechers}, {Daddi}, \& {Lentati}}]{Hodge12}
{Hodge}, J.~A., {Carilli}, C.~L., {Walter}, F., {de Blok}, W.~J.~G.,
  {Riechers}, D., {Daddi}, E., \& {Lentati}, L. 2012, \apj, 760, 11

\bibitem[{{Jablonka} {et~al.}(2013){Jablonka}, {Combes}, {Rines}, {Finn}, \&
  {Welch}}]{Jablonka13}
{Jablonka}, P., {Combes}, F., {Rines}, K., {Finn}, R., \& {Welch}, T. 2013,
  \aap, 557, A103

\bibitem[{{Kenney} \& {Young}(1988)}]{Kenney88}
{Kenney}, J.~D., \& {Young}, J.~S. 1988, \apjs, 66, 261

\bibitem[{{Kenney} \& {Young}(1989)}]{KenneyYoung89}
{Kenney}, J.~D.~P., \& {Young}, J.~S. 1989, \apj, 344, 171

\bibitem[{{Kenney} {et~al.}(1990){Kenney}, {Young}, {Hasegawa}, \&
  {Nakai}}]{Kenney90}
{Kenney}, J.~D.~P., {Young}, J.~S., {Hasegawa}, T., \& {Nakai}, N. 1990, \apj,
  353, 460

\bibitem[{{Kere{\v s}} {et~al.}(2005){Kere{\v s}}, {Katz}, {Weinberg}, \&
  {Dav{\'e}}}]{Keres05}
{Kere{\v s}}, D., {Katz}, N., {Weinberg}, D.~H., \& {Dav{\'e}}, R. 2005,
  \mnras, 363, 2

\bibitem[{{Koopmann} \& {Kenney}(2004)}]{Koopmann04}
{Koopmann}, R.~A., \& {Kenney}, J.~D.~P. 2004, \apj, 613, 866

\bibitem[{{Lee} {et~al.}(2017){Lee}, {Chung}, {Tonnesen}, {Kenney}, {Wong},
  {Vollmer}, {Petitpas}, {Crowl}, \& {van Gorkom}}]{Lee17}
{Lee}, B., {et~al.} 2017, \mnras, 466, 1382

\bibitem[{{Leroy} {et~al.}(2008){Leroy}, {Walter}, {Brinks}, {Bigiel}, {de
  Blok}, {Madore}, \& {Thornley}}]{Leroy08}
{Leroy}, A.~K., {Walter}, F., {Brinks}, E., {Bigiel}, F., {de Blok}, W.~J.~G.,
  {Madore}, B., \& {Thornley}, M.~D. 2008, \aj, 136, 2782

\bibitem[{{McDonald} {et~al.}(2009){McDonald}, {Courteau}, \&
  {Tully}}]{McDonald09}
{McDonald}, M., {Courteau}, S., \& {Tully}, R.~B. 2009, \mnras, 394, 2022

\bibitem[{{McMullin} {et~al.}(2007){McMullin}, {Waters}, {Schiebel}, {Young},
  \& {Golap}}]{McMullin07}
{McMullin}, J.~P., {Waters}, B., {Schiebel}, D., {Young}, W., \& {Golap}, K.
  2007, in Astronomical Society of the Pacific Conference Series, Vol. 376,
  Astronomical Data Analysis Software and Systems XVI, ed. R.~A. {Shaw},
  F.~{Hill}, \& D.~J. {Bell}, 127

\bibitem[{{Mok} {et~al.}(2017){Mok}, {Wilson}, {Knapen}, {S{\'a}nchez-Gallego},
  {Brinks}, \& {Rosolowsky}}]{Mok17}
{Mok}, A., {Wilson}, C.~D., {Knapen}, J.~H., {S{\'a}nchez-Gallego}, J.~R.,
  {Brinks}, E., \& {Rosolowsky}, E. 2017, \mnras, 467, 4282

\bibitem[{{Moretti} {et~al.}(2018){Moretti}, {Paladino}, {Poggianti},
  {D'Onofrio}, {Bettoni}, {Gullieuszik}, {Jaff{\'e}}, {Vulcani}, {Fasano},
  {Fritz}, \& {Torstensson}}]{Moretti18}
{Moretti}, A., {et~al.} 2018, \mnras

\bibitem[{{Muzzin} {et~al.}(2013){Muzzin}, {Wilson}, {Demarco}, {Lidman},
  {Nantais}, {Hoekstra}, {Yee}, \& {Rettura}}]{Muzzin13}
{Muzzin}, A., {Wilson}, G., {Demarco}, R., {Lidman}, C., {Nantais}, J.,
  {Hoekstra}, H., {Yee}, H.~K.~C., \& {Rettura}, A. 2013, \apj, 767, 39

\bibitem[{{Muzzin} {et~al.}(2009){Muzzin}, {Wilson}, {Yee}, {Hoekstra},
  {Gilbank}, {Surace}, {Lacy}, {Blindert}, {Majumdar}, {Demarco}, {Gardner},
  {Gladders}, \& {Lonsdale}}]{Muzzin09}
{Muzzin}, A., {et~al.} 2009, \apj, 698, 1934

\bibitem[{{Nantais} {et~al.}(2016){Nantais}, {van der Burg}, {Lidman},
  {Demarco}, {Noble}, {Wilson}, {Muzzin}, {Foltz}, {DeGroot}, \&
  {Cooper}}]{Nantais16}
{Nantais}, J.~B., {et~al.} 2016, \aap, 592, A161

\bibitem[{{Nantais} {et~al.}(2017){Nantais}, {Muzzin}, {van der Burg},
  {Wilson}, {Lidman}, {Foltz}, {DeGroot}, {Noble}, {Cooper}, \&
  {Demarco}}]{Nantais17}
---. 2017, \mnras, 465, L104

\bibitem[{{Narayanan} {et~al.}(2012){Narayanan}, {Krumholz}, {Ostriker}, \&
  {Hernquist}}]{Narayanan12}
{Narayanan}, D., {Krumholz}, M.~R., {Ostriker}, E.~C., \& {Hernquist}, L. 2012,
  \mnras, 421, 3127

\bibitem[{{Noble} {et~al.}(2017){Noble}, {McDonald}, {Muzzin}, {Nantais},
  {Rudnick}, {van Kampen}, {Webb}, {Wilson}, {Yee}, {Boone}, {Cooper},
  {DeGroot}, {Delahaye}, {Demarco}, {Foltz}, {Hayden}, {Lidman},
  {Manilla-Robles}, \& {Perlmutter}}]{Noble17}
{Noble}, A.~G., {et~al.} 2017, \apjl, 842, L21

\bibitem[{{Noeske} {et~al.}(2007){Noeske}, {Weiner}, {Faber}, {Papovich},
  {Koo}, {Somerville}, {Bundy}, {Conselice}, {Newman}, {Schiminovich}, {Le
  Floc'h}, {Coil}, {Rieke}, {Lotz}, {Primack}, {Barmby}, {Cooper}, {Davis},
  {Ellis}, {Fazio}, {Guhathakurta}, {Huang}, {Kassin}, {Martin}, {Phillips},
  {Rich}, {Small}, {Willmer}, \& {Wilson}}]{Noeske07}
{Noeske}, K.~G., {et~al.} 2007, \apjl, 660, L43

\bibitem[{{Papovich} {et~al.}(2010){Papovich}, {Momcheva}, {Willmer},
  {Finkelstein}, {Finkelstein}, {Tran}, {Brodwin}, {Dunlop}, {Farrah}, {Khan},
  {Lotz}, {McCarthy}, {McLure}, {Rieke}, {Rudnick}, {Sivanandam}, {Pacaud}, \&
  {Pierre}}]{Papovich10}
{Papovich}, C., {et~al.} 2010, \apj, 716, 1503

\bibitem[{{Papovich} {et~al.}(2016){Papovich}, {Labb{\'e}}, {Glazebrook},
  {Quadri}, {Bekiaris}, {Dickinson}, {Finkelstein}, {Fisher}, {Inami},
  {Livermore}, {Spitler}, {Straatman}, \& {Tran}}]{Papovich16}
---. 2016, Nature Astronomy, 1, 0003

\bibitem[{{Pappalardo} {et~al.}(2012){Pappalardo}, {Bianchi}, {Corbelli},
  {Giovanardi}, {Hunt}, {Bendo}, {Boselli}, {Cortese}, {Magrini}, {Zibetti},
  {di Serego Alighieri}, {Davies}, {Baes}, {Ciesla}, {Clemens}, {De Looze},
  {Fritz}, {Grossi}, {Pohlen}, {Smith}, {Verstappen}, \&
  {Vlahakis}}]{Pappalardo12}
{Pappalardo}, C., {et~al.} 2012, \aap, 545, A75

\bibitem[{{Peng} {et~al.}(2002){Peng}, {Ho}, {Impey}, \& {Rix}}]{Peng02}
{Peng}, C.~Y., {Ho}, L.~C., {Impey}, C.~D., \& {Rix}, H.-W. 2002, \aj, 124, 266

\bibitem[{{Poggianti} {et~al.}(2016){Poggianti}, {Fasano}, {Omizzolo},
  {Gullieuszik}, {Bettoni}, {Moretti}, {Paccagnella}, {Jaff{\'e}}, {Vulcani},
  {Fritz}, {Couch}, \& {D'Onofrio}}]{Poggianti16}
{Poggianti}, B.~M., {et~al.} 2016, \aj, 151, 78

\bibitem[{{Rawle} {et~al.}(2014){Rawle}, {Egami}, {Bussmann}, {Gurwell},
  {Ivison}, {Boone}, {Combes}, {Danielson}, {Rex}, {Richard}, {Smail},
  {Swinbank}, {Altieri}, {Blain}, {Clement}, {Dessauges-Zavadsky}, {Edge},
  {Fazio}, {Jones}, {Kneib}, {Omont}, {P{\'e}rez-Gonz{\'a}lez}, {Schaerer},
  {Valtchanov}, {van der Werf}, {Walth}, {Zamojski}, \& {Zemcov}}]{Rawle14}
{Rawle}, T.~D., {et~al.} 2014, \apj, 783, 59

\bibitem[{{Regan} {et~al.}(2001){Regan}, {Thornley}, {Helfer}, {Sheth}, {Wong},
  {Vogel}, {Blitz}, \& {Bock}}]{Regan01}
{Regan}, M.~W., {Thornley}, M.~D., {Helfer}, T.~T., {Sheth}, K., {Wong}, T.,
  {Vogel}, S.~N., {Blitz}, L., \& {Bock}, D.~C.-J. 2001, \apj, 561, 218

\bibitem[{{Riechers} {et~al.}(2008){Riechers}, {Walter}, {Brewer}, {Carilli},
  {Lewis}, {Bertoldi}, \& {Cox}}]{Riechers08}
{Riechers}, D.~A., {Walter}, F., {Brewer}, B.~J., {Carilli}, C.~L., {Lewis},
  G.~F., {Bertoldi}, F., \& {Cox}, P. 2008, \apj, 686, 851

\bibitem[{{Rudnick} {et~al.}(2017){Rudnick}, {Hodge}, {Walter}, {Momcheva},
  {Tran}, {Papovich}, {da Cunha}, {Decarli}, {Saintonge}, {Willmer}, {Lotz}, \&
  {Lentati}}]{Rudnick17}
{Rudnick}, G., {et~al.} 2017, \apj, 849, 27

\bibitem[{{Russell} {et~al.}(2017){Russell}, {McDonald}, {McNamara}, {Fabian},
  {Nulsen}, {Bayliss}, {Benson}, {Brodwin}, {Carlstrom}, {Edge},
  {Hlavacek-Larrondo}, {Marrone}, {Reichardt}, \& {Vieira}}]{Russell17}
{Russell}, H.~R., {et~al.} 2017, \apj, 836, 130

\bibitem[{{Scott} {et~al.}(2013){Scott}, {Usero}, {Brinks}, {Boselli},
  {Cortese}, \& {Bravo-Alfaro}}]{Scott13}
{Scott}, T.~C., {Usero}, A., {Brinks}, E., {Boselli}, A., {Cortese}, L., \&
  {Bravo-Alfaro}, H. 2013, \mnras, 429, 221

\bibitem[{{Sharda} {et~al.}(2018){Sharda}, {Federrath}, {da Cunha}, {Swinbank},
  \& {Dye}}]{Sharda18}
{Sharda}, P., {Federrath}, C., {da Cunha}, E., {Swinbank}, A.~M., \& {Dye}, S.
  2018, \mnras, 477, 4380

\bibitem[{{Sharon} {et~al.}(2013){Sharon}, {Baker}, {Harris}, \&
  {Thomson}}]{Sharon13}
{Sharon}, C.~E., {Baker}, A.~J., {Harris}, A.~I., \& {Thomson}, A.~P. 2013,
  \apj, 765, 6

\bibitem[{{Sheen} {et~al.}(2017){Sheen}, {Smith}, {Jaff{\'e}}, {Kim}, {Yi},
  {Duc}, {Nantais}, {Candlish}, {Demarco}, \& {Treister}}]{Sheen17}
{Sheen}, Y.-K., {et~al.} 2017, \apjl, 840, L7

\bibitem[{{Sofue} \& {Rubin}(2001)}]{Sofue01}
{Sofue}, Y., \& {Rubin}, V. 2001, \araa, 39, 137

\bibitem[{{Solomon} \& {Barrett}(1991)}]{Solomon91}
{Solomon}, P.~M., \& {Barrett}, J.~W. 1991, in IAU Symposium, Vol. 146,
  Dynamics of Galaxies and Their Molecular Cloud Distributions, ed. F.~{Combes}
  \& F.~{Casoli}, 235

\bibitem[{{Solomon} \& {Vanden Bout}(2005)}]{Solomon05}
{Solomon}, P.~M., \& {Vanden Bout}, P.~A. 2005, \araa, 43, 677

\bibitem[{{Stach} {et~al.}(2017){Stach}, {Swinbank}, {Smail}, {Hilton},
  {Simpson}, \& {Cooke}}]{Stach17}
{Stach}, S.~M., {Swinbank}, A.~M., {Smail}, I., {Hilton}, M., {Simpson}, J.~M.,
  \& {Cooke}, E.~A. 2017, \apj, 849, 154

\bibitem[{{Stark} {et~al.}(1986){Stark}, {Knapp}, {Bally}, {Wilson}, {Penzias},
  \& {Rowe}}]{Stark86}
{Stark}, A.~A., {Knapp}, G.~R., {Bally}, J., {Wilson}, R.~W., {Penzias}, A.~A.,
  \& {Rowe}, H.~E. 1986, \apj, 310, 660

\bibitem[{{Tacconi} {et~al.}(2010){Tacconi}, {Genzel}, {Neri}, {Cox}, {Cooper},
  {Shapiro}, {Bolatto}, {Bouch{\'e}}, {Bournaud}, {Burkert}, {Combes},
  {Comerford}, {Davis}, {Schreiber}, {Garcia-Burillo}, {Gracia-Carpio}, {Lutz},
  {Naab}, {Omont}, {Shapley}, {Sternberg}, \& {Weiner}}]{Tacconi10}
{Tacconi}, L.~J., {et~al.} 2010, \nat, 463, 781

\bibitem[{{Tacconi} {et~al.}(2013){Tacconi}, {Neri}, {Genzel}, {Combes},
  {Bolatto}, {Cooper}, {Wuyts}, {Bournaud}, {Burkert}, {Comerford}, {Cox},
  {Davis}, {F{\"o}rster Schreiber}, {Garc{\'{\i}}a-Burillo}, {Gracia-Carpio},
  {Lutz}, {Naab}, {Newman}, {Omont}, {Saintonge}, {Shapiro Griffin}, {Shapley},
  {Sternberg}, \& {Weiner}}]{Tacconi13}
---. 2013, \apj, 768, 74

\bibitem[{{Tacconi} {et~al.}(2018){Tacconi}, {Genzel}, {Saintonge}, {Combes},
  {Garc{\'{\i}}a-Burillo}, {Neri}, {Bolatto}, {Contini}, {F{\"o}rster
  Schreiber}, {Lilly}, {Lutz}, {Wuyts}, {Accurso}, {Boissier}, {Boone},
  {Bouch{\'e}}, {Bournaud}, {Burkert}, {Carollo}, {Cooper}, {Cox}, {Feruglio},
  {Freundlich}, {Herrera-Camus}, {Juneau}, {Lippa}, {Naab}, {Renzini},
  {Salome}, {Sternberg}, {Tadaki}, {{\"U}bler}, {Walter}, {Weiner}, \&
  {Weiss}}]{Tacconi18}
---. 2018, \apj, 853, 179

\bibitem[{{Tran} {et~al.}(2010){Tran}, {Papovich}, {Saintonge}, {Brodwin},
  {Dunlop}, {Farrah}, {Finkelstein}, {Finkelstein}, {Lotz}, {McLure},
  {Momcheva}, \& {Willmer}}]{Tran10}
{Tran}, K.-V.~H., {et~al.} 2010, \apjl, 719, L126

\bibitem[{{Vollmer} {et~al.}(2008){Vollmer}, {Braine}, {Pappalardo}, \&
  {Hily-Blant}}]{Vollmer08}
{Vollmer}, B., {Braine}, J., {Pappalardo}, C., \& {Hily-Blant}, P. 2008, \aap,
  491, 455

\bibitem[{{Webb} {et~al.}(2017){Webb}, {Lowenthal}, {Yun}, {Noble}, {Muzzin},
  {Wilson}, {Yee}, {Cybulski}, {Aretxaga}, \& {Hughes}}]{Webb17}
{Webb}, T.~M.~A., {et~al.} 2017, \apjl, 844, L17

\bibitem[{{Wilson} {et~al.}(2009){Wilson}, {Muzzin}, {Yee}, {Lacy}, {Surace},
  {Gilbank}, {Blindert}, {Hoekstra}, {Majumdar}, {Demarco}, {Gardner},
  {Gladders}, \& {Lonsdale}}]{Wilson09}
{Wilson}, G., {et~al.} 2009, \apj, 698, 1943

\bibitem[{{Wisnioski} {et~al.}(2015){Wisnioski}, {F{\"o}rster Schreiber},
  {Wuyts}, {Wuyts}, {Bandara}, {Wilman}, {Genzel}, {Bender}, {Davies},
  {Fossati}, {Lang}, {Mendel}, {Beifiori}, {Brammer}, {Chan}, {Fabricius},
  {Fudamoto}, {Kulkarni}, {Kurk}, {Lutz}, {Nelson}, {Momcheva}, {Rosario},
  {Saglia}, {Seitz}, {Tacconi}, \& {van Dokkum}}]{Wisnioski15}
{Wisnioski}, E., {et~al.} 2015, \apj, 799, 209

\bibitem[{{Young} {et~al.}(1995){Young}, {Xie}, {Tacconi}, {Knezek}, {Viscuso},
  {Tacconi-Garman}, {Scoville}, {Schneider}, {Schloerb}, {Lord}, {Lesser},
  {Kenney}, {Huang}, {Devereux}, {Claussen}, {Case}, {Carpenter}, {Berry}, \&
  {Allen}}]{Young95}
{Young}, J.~S., {et~al.} 1995, \apjs, 98, 219

\end{thebibliography}
\bibliographystyle{apj}

\end{document}